\documentclass[preprint2, twocolumn]{aastex6}%{emulateapj}

\usepackage{graphicx}
\usepackage{amsmath}
\usepackage{hyperref}

\usepackage{natbib}
\usepackage{xcolor}

\newcommand{\ns}{\ensuremath{n_S}}
\newcommand{\hatns}{\ensuremath{\hat{n}_S}}
\newcommand{\logp}{\ensuremath{\log_{10}\mathrm{p}}}
\newcommand{\pval}{\ensuremath{-\logp}}
\newcommand{\flux}{\ensuremath{\:\mathrm{TeV\,cm^{-2}\,s^{-1}}}}
\newcommand{\sflux}{\ensuremath{\:\mathrm{TeV^2\,cm^{-2}\,s^{-1}}}}
\newcommand{\OES}{1ES~1959+650}
\newcommand{\PKS}{PKS~1406-076}
\newcommand{\HESS}{HESS~J1616-508}
\newcommand{\UL}{$90\%$~upper-limit}
\newcommand{\dpde}{\ensuremath{d\phi/dE_\nu}}
\newcommand{\Edpde}{\ensuremath{E_\nu^2\dpde}}

\renewcommand{\deg}{\ensuremath{^\circ}}

\renewcommand{\equationautorefname}{Eq.}
\renewcommand{\figureautorefname}{Fig.}
\renewcommand{\tableautorefname}{Tab.}

\newcommand{\Autoref}[1]{%
    \begingroup%
        \renewcommand\equationautorefname{Equation}%
        \renewcommand\figureautorefname{Figure}%
        \renewcommand\tableautorefname{Table}%
        \autoref{#1}%
    \endgroup%
}

\begin{document}

\journalinfo{Version 2.0, dated \today}

\title{All-sky Search for Time-integrated Neutrino Emission from Astrophysical
       Sources with 7~yr of IceCube Data}
\shorttitle{All-sky neutrino point source searches in IceCube}
\shortauthors{M.~G.~Aartsen et al.}
\keywords{astroparticle physics --- galaxies: active --- neutrinos}
\submitted{September 16, 2016}
\accepted{December 6, 2016}

\author{
IceCube Collaboration:
\footnotesize M.~G.~Aartsen\altaffilmark{1},
\footnotesize K.~Abraham\altaffilmark{2},
\footnotesize M.~Ackermann\altaffilmark{3},
\footnotesize J.~Adams\altaffilmark{4},
\footnotesize J.~A.~Aguilar\altaffilmark{5},
\footnotesize M.~Ahlers\altaffilmark{6},
\footnotesize M.~Ahrens\altaffilmark{7},
\footnotesize D.~Altmann\altaffilmark{8},
\footnotesize K.~Andeen\altaffilmark{9},
\footnotesize T.~Anderson\altaffilmark{10}
% ...
}
\altaffiltext{1}{Department of Physics, University of Adelaide, Adelaide, 5005, Australia}
\altaffiltext{2}{Physik-department, Technische Universit\"at M\"unchen, D-85748 Garching, Germany}
\altaffiltext{3}{DESY, D-15735 Zeuthen, Germany}
\altaffiltext{4}{Dept.~of Physics and Astronomy, University of Canterbury, Private Bag 4800, Christchurch, New Zealand}
\altaffiltext{5}{Universit\'e Libre de Bruxelles, Science Faculty CP230, B-1050 Brussels, Belgium}
\altaffiltext{6}{Dept.~of Physics and Wisconsin IceCube Particle Astrophysics Center, University of Wisconsin, Madison, WI 53706, USA}
\altaffiltext{7}{Oskar Klein Centre and Dept.~of Physics, Stockholm University, SE-10691 Stockholm, Sweden}
\altaffiltext{8}{Erlangen Centre for Astroparticle Physics, Friedrich-Alexander-Universit\"at Erlangen-N\"urnberg, D-91058 Erlangen, Germany}
\altaffiltext{9}{Department of Physics, Marquette University, Milwaukee, WI 53201, USA}
\altaffiltext{10}{Dept.~of Physics, Pennsylvania State University, University Park, PA 16802, USA}
% ...
\altaffiltext{16}{Dept.~of Physics, University of California, Berkeley, CA 94720, USA}
% ...

\begin{abstract}
Since the recent detection of an astrophysical flux of high-energy neutrinos,
the question of its origin has not yet fully been answered. Much of what is
known about this flux comes from a small event sample of high neutrino purity,
good energy resolution, but large angular uncertainties. In searches for
point-like sources, on the other hand, the best performance is given by using
large statistics and good angular reconstructions. Track-like muon events
produced in neutrino interactions satisfy these requirements. We present here
the results of searches for point-like sources with neutrinos using data
acquired by the IceCube detector over 7~yr from 2008 to 2015. The discovery
potential of the analysis in the northern sky is now significantly below
$\Edpde=10^{-12}\:\mathrm{TeV\,cm^{-2}\,s^{-1}}$, on average $38\%$ lower than
the sensitivity of the previously published analysis of 4~yr exposure. No
significant clustering of neutrinos above background expectation was observed,
and implications for prominent neutrino source candidates are discussed.
\end{abstract}

\maketitle

\section{Introduction}
\label{sec:intro}

One outstanding question in astroparticle physics is the origin of
ultra-high-energy cosmic rays (UHECRs). In the paradigm of multi-messenger
astronomy, both photons and neutrinos can help resolve the sources of
UHECRs~\citep{Beatty:2009zz, Kotera:2011cp}. Photons and neutrinos are believed
to be produced in the astrophysical beam dump of cosmic-ray particles
interacting with matter at the source location. Due to lack of electric charge,
they point back to their origin, whereas cosmic rays are deflected by tangled
magnetic fields in the universe. Sources of high-energy $\gamma$-rays in our
Galaxy and extragalactic objects are detected over a wide range of
energies~\citep{Hinton:2009zz}, but both hadronic and leptonic processes
can produce $\gamma$-rays. Neutrinos, on the other hand, trace hadronic
interactions and therefore are a smoking-gun signature of cosmic-ray
acceleration~\citep{Learned:2000abc, Halzen:2002abc, Anchordoqui:2010abc,
Anchordoqui:2013dnh}. Other possible types of sources could be hidden in
$\gamma$-rays and only identified using neutrinos~\citep{Murase:2015xka,
Senno:2015tsn}.

IceCube recently reported the first observation of high-energy astrophysical
neutrinos with more than $5\sigma$ significance \citep{Aartsen:2013jdh,
Aartsen:2014gkd}.  Neutrinos with energies up to and exceeding
$1\:\mathrm{PeV}$ are observed in events starting inside the detector. Since
then, this result is confirmed in other detection channels of interactions of
$\nu_\mu+\bar{\nu}_\mu$ in the northern sky~\citep{Aartsen:2015rwa,
Aartsen:2016xlq}. The overall flux observed is consistent so far with an
isotropic emission over the full sky and all neutrino
flavors~\citep{Aartsen:2015ivb, Aartsen:2015knd}

This paper presents the most recent results of searches for point-like steady
emission of neutrinos using track-like events traversing the IceCube detector.
The statistics are increased by adding 3~yr of exposure to the previous
analyses~\citep{Abbasi:2010rd, Aartsen:2013uuv, Aartsen:2014cva}.  A sample of
$712,830$ events is obtained during 7~yr of data recording through 2015 June.
In addition, starting tracks are used in a separate sample to help access lower
energies in the southern sky~\citep{Aartsen:2016tpb}.

In \autoref{sec:icecube}, the IceCube Neutrino Observatory is introduced and
the two samples of through-going and starting tracks are characterized. In
\autoref{sec:methods}, the statistical method of unbinned likelihood
maximization for clustering searches is discussed. \Autoref{sec:results}
presents the results and their implications regarding neutrino sources, and in
\autoref{sec:conclusion} conclusions are drawn.

\section{The IceCube Neutrino Observatory}
\label{sec:icecube}

Interactions of neutrinos in IceCube are detected using Cerenkov light
emitted by relativistic charged secondary particles. To serve this purpose, one
cubic kilometer of Antarctic ice was instrumented at the South
Pole~\citep{Achterberg:2006md}. A total of 5160 digital optical modules (DOMs)
detect light emission in the ice at a depth ranging from 1450 to $2450\:$m. The
DOMs consist of a photomultiplier tube, electronics for digitization, and LEDs
for detector calibration \citep{Abbasi:2008aa, Abbasi:2010vc}.  DOMs are
attached to 86 cables (called \emph{strings}) in groups of 60 with vertical
spacing of $17\:$m; the mean distance between neighboring strings is
$\sim125\:\mathrm{m}$.  The eight innermost strings form a denser sub-array
called \emph{DeepCore} which targets lower energies~\citep{Abbasi:2011ym}. The
South Pole IceCube Neutrino Observatory includes the surface array called
\emph{IceTop}~\citep{Abbasi:2012nn} that detects and reconstructs air showers
above $300\:$TeV using 82 ice tanks. In the analysis presented here, IceTop's
capabilities are used to veto cosmic-ray-induced backgrounds.

\subsection{Neutrino detection channels}

Three event topologies are taken
into account when considering neutrino interactions in IceCube. \emph{Tracks}
are induced by muons traversing the detector. Below $700\:$GeV, muons lose
energy mainly due to ionization; above $700\:$GeV, stochastic energy losses due
to radiative emission become the dominant component. At TeV energies, muons
travel long distances, larger than several kilometers in the Antarctic
ice~\citep{Chirkin:2004hz}. Light is constantly emitted along the track. The
resulting long lever arm gives a good reconstruction performance with median
angular resolution $\Delta\Psi<1\deg$i.\footnote{Evaluated using dedicated Monte
Carlo simulation, verified using experimental data of cosmic-ray shadowing of
the Moon~\citep{Aartsen:2013zka}.} Moreover, the event rate greatly increases
because neutrinos can interact far outside the detector prior to the detection
of the secondary muon with IceCube.  Charged-current interactions of electron
or tau neutrinos, as well as neutral current interactions of any neutrino type,
produce \emph{shower}- or \emph{cascade}-like events. These types of
interactions produce almost spherically symmetric light emission, giving a
median angular resolution of $\sim$10$\deg$. Another topology is induced by
very high energy charged-current $\nu_\tau+\bar{\nu}_\tau$ interactions with
the tau lepton decaying to hadrons after traveling a distance
$\sim50\:\mathrm{m\,PeV}^{-1}\times E_\tau$, resulting in two cascades separated
distinctly. Such a \emph{double-bang} has not been observed so
far~\citep{Aartsen:2015dlt}.

Track-like events are more suited than cascades to search for very localized
(point-like), faint sources using high statistics and good angular resolution.
Such events originate primarily in neutrino charged-current interactions
of muon (anti)neutrinos with nucleons, but also in similar interactions of tau
neutrinos with the tau lepton decaying to muons and neutrinos, or
interactions of electron antineutrinos with electrons by resonant s--channel
$W^-$ exchange~\citep[Glashow Resonance;][]{Glashow:1960zz}. The energy of
a muon track is restricted to the fraction of the track visible in the
detector, limiting the energy estimation to the \emph{deposited energy} in
the detector or the muon energy upon entering the detector. Independent of the
details of neutrino production, the incident flux of astrophysical neutrinos at
Earth will consist of an approximately equally shared flavor ratio for all
neutrinos due to very long baseline neutrino oscillations~\citep{Athar:2005wg}.
In the following, only the component of tracks created in muon neutrino
interactions is taken into account as the signal of astrophysical neutrinos,
likewise to \cite{Aartsen:2014cva} and \cite{Adrian-Martinez:2014wzf}. The
impact of track events originating from other neutrino interactions is
discussed in \autoref{sec:sys}.

Construction of IceCube finished in December~2010 after 6~yr of
deployment. During construction, partial configurations of the detector were
successfully taking data, commonly indicated by IC$XY$, with $XY$ denoting the
number of active strings. The first three years of the event sample are in
partial detector configurations IC40, IC59, and IC79. The previously published
analysis~\citep{Aartsen:2014cva} included the first year of data-taking with
the completed detector IC86 and 3~yr in partial
configuration~\citep{Abbasi:2010rd, Aartsen:2013uuv}. These samples focus on
through-going track-like events, yielding high statistics over broad energy
ranges. In the southern sky, an additional selection of starting tracks, which
has a greatly reduced background rate, is performed access lower
energies~\citep{Aartsen:2016tpb}.  This uses completely independent events to
the ones selected in the aforementioned through-going track channel.

\begin{deluxetable}{p{0.08\textwidth}rrrr}
\tablecaption{IceCube samples used in this analysis. For each sample,
              characteristic features are quoted, separated in the two halves
              of the sky. Previously published results used four years of data:
              IC40~\citep{Abbasi:2010rd}, IC59+IC79~\citep{Aartsen:2013uuv},
              and the first year of IC86~\citep{Aartsen:2014cva}. Data taken in
              the seasons from 2012 to 2015 are added in this paper. The
              separation in the two regions is done at a declination of
              $\delta=-5\deg$. In addition, starting tracks are used in a
              separate sample~\citep{Aartsen:2016tpb} with two additional years
              of available data.\label{tab:sample-stats}}

\tablehead{\colhead{Sample} & \colhead{Livetime} & \colhead{atm. $\nu$}
            & \colhead{Up-going} & \colhead{Down-going}\\
           & \colhead{(days)} & \colhead{(day$^{-1}$)}
            & \colhead{events} & \colhead{events}}
\startdata
IC40          & 376   &  40         &  16323 &  20577 \\*
IC59          & 348   & 120         &  48105 &  58906 \\*
IC79          & 316   & 180         &  54823 &  38310 \\*
IC86          & 333   & 210         &  67938 &  68302 \\*
2012-2015     & 1058  & 220         & 235602 & 102983 \\*
$\Sigma$      & 2431 & \nodata      & 422791 & 289078 \\*
starting tracks & 1715  & $<\,0.03$ &      0 &    961 \\*
\enddata
\end{deluxetable}

The details of all samples are listed in \autoref{tab:sample-stats}, including
the exposure time and sample size. In this work, the detector livetime is
increased by adding data from June~2012 to June~2015 to the analysis, thus
increasing the livetime by 1058~days, to a total of 2431~days. The sample of
starting tracks~\citep{Aartsen:2016tpb} has an increased livetime of 2~yr to
a total of 1715~days that coincide with the five most recent years of the
through-going track event sample, starting with IC79.

In the following, the modeling of signal and background for both the
through-going and starting track samples is described.  The event selection is
briefly discussed and the performance of the event sample highlighted. For more
detailed information, refer to \cite{Abbasi:2010rd}, \cite{Aartsen:2013uuv},
and \cite{Aartsen:2014cva} for through-going muons and \cite{Aartsen:2016tpb}
for starting tracks.

In this work, muon tracks induced by astrophysical neutrino interactions are
the main signal category in the search for point-like sources of neutrinos.
Detailed Monte Carlo simulation is used to evaluate the response of IceCube to
such events and distinguish them from atmospheric backgrounds. The median
angular resolution $\Delta\Psi$ and the event rate expectation
\begin{align}
    \dot{N}_\nu=\int d\Omega\int_0^\infty dE\,A_\mathrm{eff}\left(E, \Omega\right)
                \times F_\nu\left(E_\nu,\Omega\right)
    \label{eq:ev_rate}
\end{align}
given by the detector \emph{effective
area} $A_\mathrm{eff}$ and the incident neutrino flux $F_\nu$ can be derived
from the simulation (for more information about the details of the Monte Carlo
simulation used here, see \cite{Aartsen:2016xlq}).

\subsection{Through-going tracks}
\label{sec:thru}

\begin{figure}[t]
    \centering
    \includegraphics[width=\columnwidth]{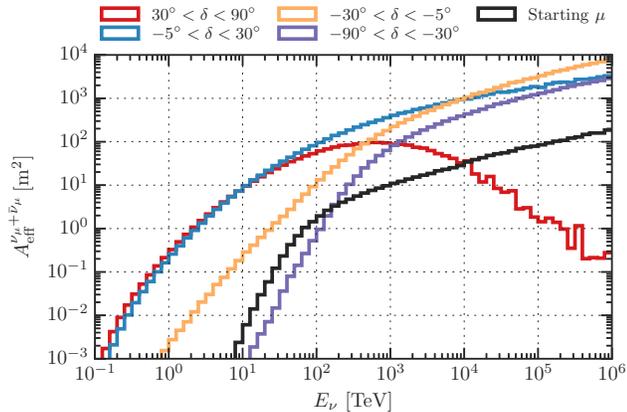}
    \caption{IceCube effective area as defined in \autoref{eq:ev_rate} versus
             neutrino energy for a flux of $\nu_\mu+\bar{\nu}_\mu$ calculated
             using simulation of neutrino events for the selection of IC86
             (seasons 2012-2015) described in \autoref{sec:thru}.  The
             effective area for through-going muons is averaged over the solid
             angle in the declination range ($\delta$) indicated in the legend.
             Additionally, the effective area for starting tracks in the
             southern sky ($\delta<-5\deg$) is shown in black (cf.
             \autoref{sec:starting}).}
    \label{fig:effA}
\end{figure}

The main background regarding neutrino searches in IceCube consists of
atmospheric muons that are created in extensive air showers and reach IceCube
at a depth of $\sim2\:\mathrm{km}$.  Almost all events triggering the detector
at a rate of $\sim2.8\:\mathrm{kHz}$ belong to this component. Similar
to~\cite{Aartsen:2014cva}, the selection is split into two regions divided at
the horizon (declination $\delta=-5\deg$) due to different background
characteristics, as explained in the following.

\begin{figure}[t]
    \centering
    \includegraphics[width=\columnwidth]{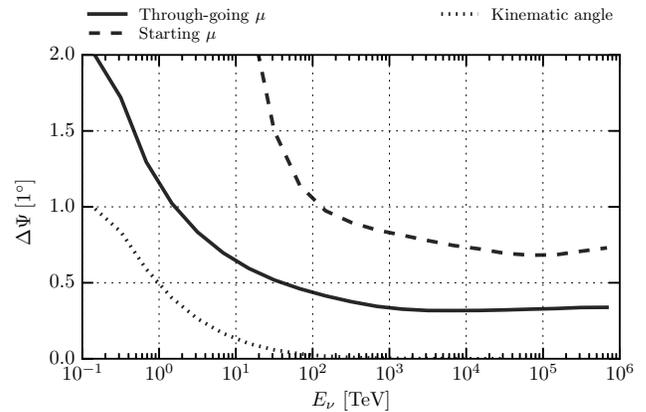}
    \caption{IceCube median angular resolution versus neutrino energy for
             $\nu_\mu+\bar{\nu}_\mu$ calculated from Monte Carlo simulation for
             the IC86 sample described in \autoref{sec:thru}. Through-going
             tracks (solid black) are shown together with starting tracks
             (dashed black; see \autoref{sec:starting}). Moreover, the
             median kinematic angle of the secondary muon in CC neutrino
             interactions is shown~(dotted black line).}
    \label{fig:mrs}
\end{figure}

In the northern sky (up-going region, $\delta\geq-5\deg$), the main background
consists of atmospheric muon events that are mis-reconstructed as up-going;
truly up-going muons can only originate from prior neutrino interactions, since
all other muons are shielded by the Earth. Hence, the atmospheric muon
background is rejected by identifying poorly reconstructed events. In order to
achieve this, multivariate selection techniques (boosted decision tree, BDT)
are used to discriminate well-reconstructed tracks from neutrino interactions
against mis-reconstructed background. The variables used in the BDT are
connected to the event quality and a clear track-like topology. Similar to
\cite{Aartsen:2014cva}, BDTs are trained for two signal energy spectra of
$E^{-2}$ and $E^{-2.7}$ and the cut on the linear combination of BDT scores is
optimized to yield the best sensitivity and discovery potential over a wide
range of energies with the final cut on BDT output. These spectra are chosen to
be sensitive to both hard energy spectra as well as soft or cutoff spectra. A
neutrino-dominated sample is obtained in the northern sky. The remaining
background consists of atmospheric neutrinos produced in the northern sky.
These neutrinos are an irreducible background, but follow a softer energy
spectrum ($\sim E^{-3.7}$--$E^{-4.0}$ or $\sim E^{-2.7}$--$E^{-3.0}$ for
conventional or prompt neutrinos, respectively) than the expected signal ($\sim
E^{-2}$ consistent with diffuse muon neutrino signal at $250\:\mathrm{TeV}$ and
above; see \cite{Aartsen:2016xlq}). \Autoref{fig:effA} shows the effective area
calculated using muon neutrino Monte Carlo simulation of the final event sample
for different declination regions.  In the northern sky (red, blue), a low
energy threshold is achieved, while for near-vertically up-going events
absorption becomes an important effect above $100\:\mathrm{TeV}$.
\Autoref{fig:mrs} shows the median angular resolution of the track
reconstruction (solid) with respect to the primary neutrino direction against
neutrino energy. Above TeV energies, the kinematic angle (dotted) of muon and
parent neutrino becomes negligible and the angular resolution is below $1\deg$.
The energy resolution of the track's energy proxy is $\sim30\%$ in
$\log_{10}{E}$~\citep{Aartsen:2013vja}. This is only a lower limit on the
energy of a muon that enters the detector from outside and loses energy prior
to detection, as well as on energy of the primary neutrino that produced the
muon in an interaction with a nucleus. For more information about the neutrino
and muon energy estimation, refer to \cite{Aartsen:2015rwa, Aartsen:2016xlq}.

\begin{figure}[t]
    \centering
    \includegraphics[width=\columnwidth]{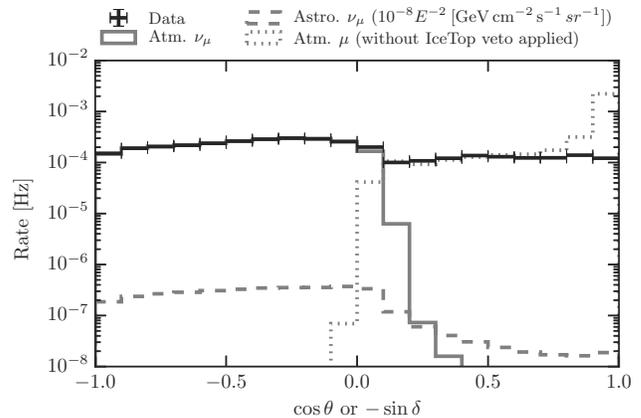}
    \caption{Zenith ($\cos\theta$) or declination ($-\sin\delta$) distribution
             of the through-going track sample after event selection (2012--2015
             data). Values of $-1$ correspond to vertically up-going events.
             Shown is the experimental data (black), compared to the
             atmospheric $\nu_\mu+\bar{\nu}_\mu$ expectation of conventional
             atmospheric (solid gray) and astrophysical neutrinos (dashed
             gray), and atmospheric muons (dotted gray) from Monte Carlo
             simulation. For simulated atmospheric muons, the plot shows the
             distribution without the IceTop veto applied.}
    \label{fig:cosSpline}
\end{figure}

In the southern sky (down-going, $\delta<-5\deg$), the picture changes because
of large backgrounds of well-reconstructed down-going atmospheric muons.
Moreover, muons are produced at high multiplicity in cosmic-ray showers,
resulting in bundles of muons; such bundles produce large amounts of light in
the detector, thus imitating the signature of a single muon of much higher
energy. Similar to the northern sky, BDTs are used to select only the
best-reconstructed events at the highest energies. Following the development of
\cite{Aartsen:2014cva}, in addition to event quality and track topology
parameters, four variables are used to further discriminate atmospheric
muon bundles from single muons. These variables use the deposited energy along
the track, as well as the light-arrival time of photons at the DOMs. The
Cerenkov light yield for high-energy muons is dominated by stochastic cascades
from energy losses along the track superimposed to the Cerenkov cone of the
muon track. Bundles of muons at lower energies show less frequent losses by
stochastic cascades. The result is a smoother light yield along the
track. Furthermore, muon bundles consist of a superposition of many
Cherenkov-cones, resulting in many photons arriving earlier than under the
assumption of one single Cherenkov cone. For a signal-spectrum $E^{-2}$, a BDT
is trained for final selection of events. The large backgrounds require harsh
cuts to reduce their rate significantly, resulting in an effective selection of
only very high energy events, as shown in \autoref{fig:effA} (yellow and
purple).  It is evident that the energy threshold in the down-going region
increases to $\sim100\:\mathrm{TeV}$ and even further for more vertically
down-going events. The IceTop surface array is used as an active veto against
coincident air-shower events for vertically down-going events.  For
high energies, this vetoes $90\%$ of the events for vertically down-going
events and less for inclined events, with random coincidences in less than
$0.1\%$ of the cases \citep{Aartsen:2013uuv}; compare \autoref{fig:cosSpline}.
The final event rate in the southern sky is optimized to yield the best
sensitivity and discovery potential for an $E^{-2}$ spectrum.

The distribution of the cosine zenith $\cos\theta$ (equivalent to the
negative sine of declination $-\sin\delta$ in equatorial coordinates) is shown
in \autoref{fig:cosSpline}. The event rate for experimental data of all 3~yr is
compared to the expectations of muon neutrinos and atmospheric mouns estimated
from Monte Carlo simulation. In the northern sky ($\theta>85\deg$), the sample
is dominated by atmospheric neutrinos produced in the decays of kaons and pions
in cosmic-ray air showers~\citep{Honda:2006qj} and is well described by Monte
Carlo simulation. In the southern sky, the neutrino event rate reduces
drastically due to the higher energy threshold mentioned before.  Instead,
atmospheric muons are the dominant component. Astrophysical neutrinos with hard
energy spectra (as, e.g. shown with $E^{-2}$) do not suffer a rate loss as
severe as for the soft energy spectrum of atmospheric neutrinos. For very
vertically down-going events, the IceTop surface array vetoes atmospheric muon
and neutrino events because of their coincident air shower. In this region, the
observed event rate is kept constant with the rest of the southern sky using
looser cuts on the BDT score, which allows more neutrinos to be detected
especially at the lower energy end. Hence, the corresponding event rate in
\autoref{fig:cosSpline} does not decrease above $\cos\theta>0.7$.

\subsection{Starting tracks}
\label{sec:starting}

In the southern sky, the large background of atmospheric muons reduces the
efficiency to select through-going tracks induced by neutrinos below the PeV
regime. A very large fraction of the aforementioned background can be rejected
by imposing an active veto at the detector boundary, as, for example, used in
\cite{Aartsen:2013jdh}. This reduces the detector volume to a smaller fraction
of the instrumented volume sacrificing statistics for signal purity.
Furthermore, the more clearly an event is identified as a starting track, the
more probable it is to be an astrophysical rather than atmospheric background.
Down-going atmospheric neutrino events at high energy are likely to be
accompanied by muons produced in the same cosmic-ray shower that triggers the
veto and reduces the atmospheric neutrino background~\citep{Schonert:2008is,
Gaisser:2014bja}. In analyses using veto techniques~\citep{Aartsen:2013jdh,
Aartsen:2014muf}, the selection is usually more efficient for cascade-like
events than tracks, and high astrophysical neutrino purity demands neglecting
energies below $60\:$TeV where backgrounds are more abundant.  In searches for
point-like sources of astrophysical neutrinos, track-like events are of great
importance given their good angular resolution compared to cascade-like events.
Furthermore, the purity demands are lower since the signal of a point-like
source is restricted to a small portion of the sky, hence reducing the
background significantly. Consequently, the minimum required total charge
deposited in the PMTs of the IceCube detector by an event is lowered to
$1500\:\mathrm{p.e.}$ compared to
$6000\:\mathrm{p.e.}$~\citep{Aartsen:2013jdh}, resulting in a higher signal
efficiency at lower energies. In addition, only down-going tracks are used, and
cuts are imposed that select well-reconstructed track-like
events~\citep{Aartsen:2016tpb}. For $\nu_\mu+\bar{\nu}_\mu$ events at energies
smaller than $200\:\mathrm{TeV}$, the effective area of the analysis is bigger
than for vertically through-going tracks
($\delta<-30\deg$; \autoref{fig:effA}).  For energies up to
$1\:$PeV, the effective area is smaller, but a higher purity is achieved.  The
angular resolution for starting tracks is shown in \autoref{fig:mrs} (dashed)
and is $\sim1\deg$ in the interesting energy region; the reconstruction is
worse than for through-going events (solid), due to a smaller lever arm for
tracks starting within the fiducial volume of the detector.

In 5~yr, 961 events were recorded in the southern sky starting track
sample. The overlap of events in the starting track sample and the
through-going events is very small, and overlapping events are removed from the
through-going sample because of its higher background rate.

\section{Methods}
\label{sec:methods}

\begin{figure}[t]
    \centering
    \includegraphics[width=\columnwidth]{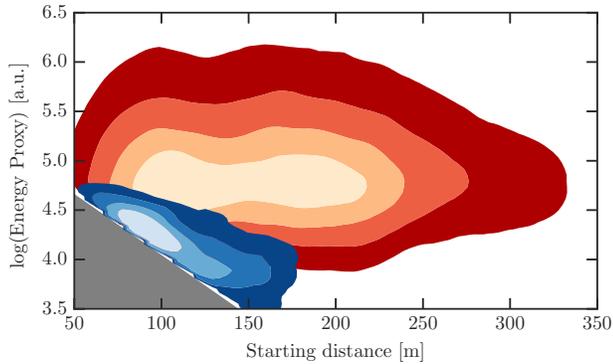}
    \caption{Probability distribution of starting distance versus
             energy proxy (logarithmic, in arbitrary units,
             a.u.) for both atmospheric background (blue) and neutrinos (red)
             with $E^{-2}$ spectrum. Different contours depict regions of
             $20\%$ coverage each. The gray shaded areas shows a region of no
             coverage resulting from a cut in the selection of
             \cite{Aartsen:2016tpb}.}
    \label{fig:mese_dist_logE}
\end{figure}

In order to look for clustering in the sky, the analysis uses an unbinned
likelihood maximization, similar to the previous
analyses~\citep{Aartsen:2014cva}. The unbinned likelihood is defined as
\begin{align}
  \label{eq:unbinnedLLH}
  \begin{split}
    \mathcal{L}\left(\ns,\gamma\right)
        =\prod_i&\left(\frac{\ns}{N}\mathcal{S}
                               \left(\left|\mathbf{x}_S-\mathbf{x}_i\right|,\,E_i;\,\gamma\right)\right.\\
                      &\quad\left.+\left(1-\frac{\ns}{N}\right)\mathcal{B}
                               \left(\sin\delta_i,\,E_i\right)
                       \right)
  \end{split}
\end{align}
using multiple observables that are introduced in the following. The signal
hypothesis used in this work assumes time-integrated emission of neutrinos.
Hence, the signature reduces to spatial clustering modeled with a two
dimensional Gaussian
$\exp\left(-\left|\mathbf{x}_S-\mathbf{x}_i\right|^2/2\sigma_i^2\right)
/\left(2\pi\sigma_i^2\right)$ using the reconstruction uncertainty $\sigma_i$
estimated on an event-by-event basis~\citep{Neunhoffer:2004ha, Abbasi:2010rd}.
The probability distribution function for the spatial distribution of
background is estimated using experimental data and depends only on the event's
declination $\delta_i$, the probability in right ascension is distributed
uniformly $1/2\pi$. This yields
$\mathcal{P}_\mathcal{B}\left(\sin\delta_i\right)/2\pi$ for the spatial
probability of background.

In addition, energy information is used to distinguish background with soft
spectra ($E^{-3.7}$) from signal with harder spectra of typically $E^{-2}$.
Hence, for each event, probability distributions
$\mathcal{E}_\mathcal{S/B}\left(E_i\right)$ for signal and background are
evaluated using the event's energy proxy $E_i$. For signal, an
unbroken power-law with variable index $\gamma$, $\dpde\propto E^{-\gamma}$ is
used, the background is estimated from experimental data. The estimation is
done declination-dependent, because of the energy-dependence of the effective
area (\Autoref{fig:effA}). This yields the final modeling of probabilities for
signal and background
\begin{align}
  \begin{split}
    \mathcal{S}=&\frac{1}{2\pi\sigma_i^2}
        \mathrm{e}^{-\frac{\left|\mathbf{x}_S-\mathbf{x}_i\right|^2}{2\sigma_i^2}}
        \times\mathcal{E}_\mathcal{S}\left(E_i, \sin\delta_i;\gamma\right)\\
    \mathcal{B}=&\frac{\mathcal{P}_{\mathcal{B}}\left(\sin\delta_i\right)}{2\pi}
        \times\mathcal{E}_\mathcal{B}\left(E_i,\sin\delta_i\right)
  \end{split}\label{eq:SB}
\end{align}
entering the unbinned likelihood calculation in \autoref{eq:unbinnedLLH}. Two
parameters are fit in the likelihood, the number of source events
$\ns\geq0$ and the source spectral index $\gamma\in\left[1,4\right]$.

In contrast to through-going tracks (\Autoref{sec:thru}), no declination
dependence of the energy distribution is observed for starting tracks
(\Autoref{sec:starting}). This is due to two reasons: the sample only uses
down-going events so Earth absorption does not occur, and a uniform charge
threshold over all declinations is applied, yielding a uniform effective
area~\citep{Aartsen:2016tpb}.  Meanwhile, starting tracks carry more
information than direction and energy; for starting tracks, the vertex of a
neutrino interaction can be reconstructed from the first visible light. A
background of atmospheric muons can sneak past the veto by not emitting enough
light prior to detection. Nevertheless, the higher the energy of the
reconstructed track gets, the more likely it is for the vertex to be
reconstructed close to the detector boundary due to constant light emission
along the track, as shown in \autoref{fig:mese_dist_logE}. A clear
anti-correlation of event energy and \emph{starting distance}\footnote{Distance
along the track pointed back from the reconstructed vertex to the entry point
in the detector, see \cite{Aartsen:2016tpb} for more information.} is
observed for background events (blue). Truly starting signal neutrinos (red) do
not show this correlation because the entering neutrino does not emit light.
Consequently, the starting distance $d_i$ can be used in addition to the event
energy, to disentangle signal and background, modifying the energy likelihood
$\mathcal{E}\left(E_i\right)\rightarrow\mathcal{E}\left(E_i,d_i\right)$,
resulting in an additional discrimination power at lower energies.

\begin{figure}[t]
    \centering
    \includegraphics[width=\columnwidth]{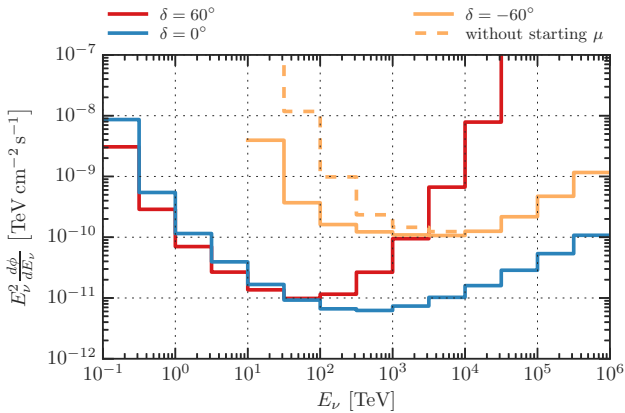}
    \caption{Discovery potential ($5\sigma$) for this analysis in different
             bins of neutrino energy $E_\nu$ with half-decade width. Within
             this energy range, an $E_\nu^{-2}$ spectrum is used. Three
             different declinations are shown: Up-going (red,
             $\delta=60\deg$), horizontal (blue, $\delta=0\deg$), and
             down-going (yellow, $\delta=-60\deg$) events. For down-going
             events, the dashed line shows the discovery potential \emph{not
             using} the starting track sample described in
             \autoref{sec:starting}.}
    \label{fig:diff-sens}
\end{figure}

As in the previous analysis~\citep{Aartsen:2014cva}, the different samples
listed in \autoref{tab:sample-stats} consist of different detector
configurations including partial detector configurations, plus samples using
only starting tracks. The total likelihood of all combined samples is the
product of all individual likelihoods, or the sum of the logarithms,
$\log\mathcal{L}\left(\ns,\gamma\right)
=\sum_j\log\mathcal{L}\left(\ns^j,\gamma\right)$ for all samples $j$. In the
scenario of steady emission, the total number of signal events $\ns$ is split
proportionally among the samples given their exposure time and expected signal
statistics derived from the effective area (\Autoref{fig:effA} and
\autoref{eq:ev_rate}) and the value of the spectral index fitting parameter
$\gamma$:
\begin{align}
    \ns^j=\ns\times
        \frac{\int\limits_0^\infty dE\,A^j_\mathrm{eff}\left(E,\sin\delta\right)\,E^{-\gamma}}
             {\sum\limits_i\int\limits_0^\infty
             dE\,A^i_\mathrm{eff}\left(E,\sin\delta\right)\,E^{-\gamma}}
    \label{eq:nsj}
\end{align}
Thus, the unbinned likelihood in \autoref{eq:unbinnedLLH} is maximized using
two parameters only for all samples, that is the number of signal-like events
$\ns$ and the spectral index $\gamma$. The null hypothesis is the observation
of no signal-like events $\ns=0$ and defines the test statistic of best-fitting
hypothesis $\left(\hatns, \hat{\gamma}\right)$ over null hypothesis,
$\mathcal{TS}=2\log\left(\mathcal{L}\left(\hatns,\hat{\gamma}\right)
/\mathcal{L}\left(\ns=0\right)\right)$. In the maximization of the test
statistic, only over-fluctuations $\ns\geq0$ are taken into account; negative
$\ns$ are not part of the physics scenario of neutrino
sources~\citep{Braun:2008bg} and not considered here. Thus, the test-statistic
of the null hypothesis is expected to split in two fractions, one bound at
$\ns=0$ and over-fluctuations $\ns>0$. The latter are distributed according to
$\chi^2_{n_\mathrm{dof}}$-statistics with $n_\mathrm{dof}\sim1.5$, less than the
number of free parameters (2). This is due to $\ns$ and $\gamma$ being partly
degenerate; moreover $\gamma$ is only defined for $\ns>0$ as can be seen in
\autoref{eq:unbinnedLLH}. The fraction $\eta$ of over-fluctuations ranges from
$50\%$ to $30\%$ in the northern and southern sky, respectively. From the
estimation of the test statistic distribution, the p-value
$\mathrm{p}=\eta\times\int_\mathcal{TS}^\infty dX\,
\chi^2\left(X\left|n_\mathrm{dof}\right.\right)$ of an observation being
consistent with background can be calculated. The p-value will mostly be quoted
as $\pval$ in the following.

\subsection{Neutrino point source sensitivity}
\label{sec:sensitivity}

\Autoref{fig:diff-sens} shows the $5\sigma$ discovery-potential of the unbinned
likelihood analysis versus neutrino energy for point sources at various
declinations.\footnote{The discovery potential is defined as a
\emph{false-positive} rate of $5\sigma$ or $2.87\times10^{-7}$ with
\emph{false-negative} of $50\%$. The sensitivity is defined as a
\emph{false-positive} rate of $50\%$ with \emph{false-negative} of $10\%$.} A
$\nu_\mu+\bar{\nu}_\mu$ neutrino signal with half-decade width is used for
signal injection, using an $E^{-2}$ spectrum within the energy range indicated
by the step function. The discovery potential shows a strong variation with
declination. In the up-going region ($\delta\geq-5\deg$), atmospheric muon
background is efficiently removed and a large effective area with good angular
resolution is achieved above TeV energies, compare \autoref{fig:effA} and
\autoref{fig:mrs}. This yields a discovery potential reaching from TeV to EeV
energies at the horizon ($\delta=0\deg$, blue). For vertically up-going events
($\delta=60\deg$, red), neutrinos at energies above $100\:\mathrm{TeV}$ begin
to be absorbed in the Earth, hence reducing the discovery potential compared to
the horizon.

In the down-going region (southern sky, $\delta<-5\deg$), large backgrounds of
atmospheric muons result in a higher energy threshold of
$\sim100\:\mathrm{TeV}$.  Moreover, muon bundles imitate single muons at very
high energies resulting in a high energy background. This diminishes the
performance compared to the northern sky. At $\delta=-60\deg$ (yellow), the
discovery potential is most effective above energies of $~100\:\mathrm{TeV}$;
in fact, in between $100\:\mathrm{TeV}$ and PeV energies, starting tracks
described in \autoref{sec:starting} dominate the sensitivity compared with
through-going muons (\Autoref{sec:thru}).  Even though the effective area is of
the same order in this energy regime, starting tracks have significantly less
background and thus a $\sim170\times$~higher purity
($0.09\:\mathrm{d}^{-1}\,\mathrm{sr}^{-1}$ background events for starting
tracks compared to $15.5\:\mathrm{d}^{-1}\,\mathrm{sr}^{-1}$ for through-going
tracks). Including starting tracks gives a factor of $\sim8$ improvement in
discovery potential at $100\:\mathrm{TeV}$ compared to only using through-going
events (dashed yellow in \autoref{fig:diff-sens}). In the southern sky, similar
searches of \cite{Adrian-Martinez:2014wzf} test much lower energies, resulting
in complementary results combined in \cite{Adrian-Martinez:2015ver}.

\subsection{Full-sky search}
\label{sec:allsky}

To find the most significant clustering in the sky, the unbinned likelihood
maximization is performed on the entire sky. This is done iteratively using a
grid with isotropically spaced points~\citep{Gorski:2004by} finer than the
typical event reconstruction uncertainty that enters the likelihood description
in \autoref{eq:SB}.

Thus, for any point in the sky, the best fitting $\hatns$,
$\hat{\gamma}$ and the test statistic $\mathcal{TS}$ are obtained. The
direction with the smallest p-value defines a \emph{hot-spot} showing the
biggest deviation from background expectation. This is done for northern and
southern sky separately, as they differ in atmospheric backgrounds and energy
reach.  Thus, two positions in the sky will be reported in the full-sky scan.
The significance is trial corrected accounting for the chance of background
fluctuations ocurring at any position in the sky. The probability to observe no
pre-trial $\logp$ that is smaller than the one at the \emph{hot-spot} for $N$
independent trials is given by
\begin{align}
  \begin{split}
    d\mathcal{P}
        %=\ln\left(10\right)\,N\,\left(1-10^{\logp}\right)^{N-1}10^{\logp}\,d\left(\logp\right)\,.
        =N\,\left(1-\mathrm{p}\right)^{N-1}\,d\mathrm{p}\,.
  \end{split}
  \label{eq:trial_corr}
\end{align}
For both northern and southern sky, the effective number of independent trials
$N$ in the sky is fitted to $\sim190,000$ by repeating the analysis on
scrambled data maps. Regions close to the poles ($5\deg$) are excluded
from the scan because no large off-source regions are available for scrambling.
Accounting for the trial factor a pre-trial significance of $5.67\sigma$
(p-value $7.13\times10^{-9}$) is needed for a \emph{hot-spot} to be detected at
$3\sigma$ significance in the scan of the full sky.

\subsection{Hotspot population analysis}
\label{sec:population}

The large trial factor of the full-sky scan requires very strong sources that
overcome the trial factor. Thus, in addition to looking at only the most
significant spot in the sky, the entire sky can be tested for an accumulation
of multiple spots at intermediate significance that exceeds the number expected
by background.

From the scan of the full sky, the positions of all the local maxima exceeding
$\pval\geq3$ are selected. The number of spots observed above a threshold of
$\pval_\mathrm{min}$ is compared to background expectation from repeating the
analysis on scrambled data maps.  The threshold value is optimized to give the
most significant excess over mean background expectation
$\lambda\left(\pval_\mathrm{min}\right)$ with Poisson statistics. The p-value
of the observation of at least $n$ spots, given this expectation, is
\begin{align}
    \mathcal{P}=\mathrm{e}^{-\lambda}\sum_{m=n}^\infty\frac{\lambda^m}{m!}
    \label{eq:HPApval}
\end{align}
and defines the test statistic of this test. Due to the optimization of the
threshold $\pval_\mathrm{min}$ to minimize $\mathcal{P}$, the final result is
trial corrected using scrambled experimental data.  This is done separately for
northern and southern sky. In addition, parts of the sky coinciding with the
Galactic plane~$\pm15\deg$ are analyzed as well for a Galactic contribution.

\subsection{Source list searches}
\label{sec:sourcelists}

In addition to the previously explained searches that did not make any prior
assumptions regarding directions in the sky, sources of high-energy
$\gamma$-rays can be used to search for neutrino emission. Thus, the trial
factor of the unbiased full-sky scan can be effectively reduced by probing 74
promising sources selected \emph{a-priori}.  The sources are collected in two
lists as used in \cite{Aartsen:2014cva} and \cite{Adrian-Martinez:2014wzf,
Aartsen:2016tpb}, respectively.  The first one (44 sources) probes mainly the
northern sky (34 candidates) and the largest fraction of sources are
extra-galactic objects. The list can be found in \autoref{tab:sourcelist1}. The
second list (\Autoref{tab:sourcelist2}) of 30 candidates focuses on the
southern sky, especially Galactic sources.

The most significant source in each list will be trial corrected given the
number of sources. For the first list, this procedure will be done separately
for northern and southern sky. Hence, three post-trial p-values are reported.

\subsection{Systematic uncertainties}
\label{sec:sys}

All analyses described in the previous sections are robust against systematic
uncertainties. Background is estimated using experimental data that is
scrambled in right ascension and does not require dedicated Monte Carlo
simulation. For the calculation of neutrino fluxes however, Monte Carlo
simulation is needed to insert neutrino sources into background maps. The
resulting neutrino fluxes are affected by systematic uncertainties. Especially,
the evaluation of the effective area and reconstruction performance is affected
by systematic effects in Monte Carlo, and thus the sensitivity to neutrino
fluxes (\Autoref{eq:ev_rate}). Main systematic uncertainties include the
optical properties (scattering, absorption) of the South Pole
ice~\citep{Aartsen:2013rt}, the optical efficiency of Cherenkov light
production yield and detection in the DOMs~\citep{Abbasi:2010vc}, and different
photo-nuclear interaction models~\citep{Bugaev:2002gy, Bugaev:2003bm,
Abramowicz:1991xz, Abramowicz:1997ms}. All systematic effects are propagated
through the entire likelihood analysis described in \autoref{sec:methods} to
obtain the uncertainties on the fluxes using $\dpde\propto E^{-2}$ spectra. The
biggest impact on the fluxes comes from varying the optical efficiency by
$\pm10\%$, resulting in a flux uncertainty of $7.5\%$. Increasing the
absorption or scattering of photons in ice by $10\%$ affects the flux by
$5.6\%$.  Uncertainties in the photo-nuclear
cross-sections~\citep{Bugaev:2002gy, Bugaev:2003bm} result in an flux
uncertainty of similar size with $5.9\%$.  Adding these values in quadrature
yields a total systematic uncertainty of $11\%$ on $\nu_\mu+\bar{\nu}_\mu$
fluxes quoted in the following.

For all locations tested, only the maximal likelihood values of $\hatns$ and
$\hat{\gamma}$ are reported. Because of small event statistics at the position
of the likelihood maximization and limited energy resolution of the neutrino
energy (compare \autoref{sec:thru}), uncertainties on the spectral index are of
the order $\pm1$ and reduce to $\pm0.5$ for values of $\ns$ of $\sim15$ and
$\sim50$, respectively~\citep{Braun:2008bg}. Hence, the impact of systematic
uncertainties in the energy reconstruction is small compared to the statistical
limitations.

Albeit not a systematic uncertainty per se, so far only fluxes of
$\nu_\mu+\bar{\nu}_\mu$ were considered. This is a conservative estimate,
because track-like events can also originate in other cases that are discussed
in the following. Firstly, tau-leptons created in charged-current
$\nu_\tau+\bar{\nu}_\tau$ interactions decay into muons with $17\%$ branching
ratio~\citep{Jeong:2010nt, Agashe:2014kda}, resulting in a muon track with
lower energy due to the three-body decay $\tau\rightarrow\mu\nu_\mu\nu_\tau$.
This decay is important for up-going events, because secondary neutrinos are
produced in $\tau$-neutrino regeneration during propagation. Secondly,
interactions of $\bar{\nu}_e+e^-\rightarrow W^-$ at the
Glashow-resonance~\citep{Glashow:1960zz} at $6.3\:\mathrm{PeV}$ produce
tracks~($\bar{\nu}_e+e^-\to\bar{\nu}_\mu+\mu^-$) at $10.6\%$ branching
ratio~\citep{Agashe:2014kda}. Lastly, at the highest energies above PeV,
$\tau$-neutrino induced double bangs are well-reconstructable and further
increase the number of $\tau$-flavored events in the sample. Accounting for
these fluxes assuming an equal flavor ratio at Earth reduces the per-flavor
flux necessary for detection by $5\%$ assuming an unbroken $E^{-2}$ spectrum.
For harder spectra, the sensitivity gain due to regeneration effects in the
northern sky becomes stronger. For example, a spectrum of $\dpde\propto E^{-1}$
has an $30\%$ improved sensitivity compared to only considering muon neutrinos.
This greatly increases the sensitivity with respect to models that predict very
hard neutrino energy spectra peaking above
PeV~energies~\citep{Petropoulou:2015upa, Reimer:2015abc}.

\section{Results and implications}
\label{sec:results}

\begin{figure}[t]
    \centering
    \includegraphics[width=\columnwidth]{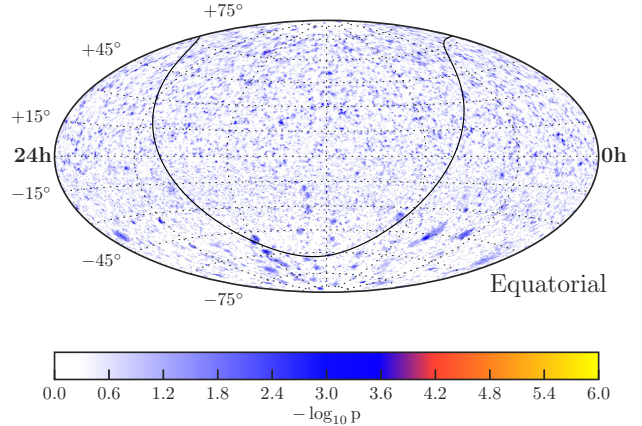}
    \caption{All-sky result of the unbinned likelihood maximization shown in
             equatorial coordinates (J2000). Shown is the negative logarithm of
             the pre-trial p-value, $\pval$, assuming no
             clustering as null-hypothesis. The Galactic Plane is shown as
             black line.}
    \label{fig:skymap}
\end{figure}

In the unbinned likelihood analysis using seven years of IceCube livetime, no
significant excess of astrophysical neutrino sources was found. In the
following, the results of the three tests introduced in the previous sections
are discussed and \UL s on neutrino source fluxes are calculated. Finally,
implications with respect to neutrino models of $\gamma$-ray sources and the
observed diffuse neutrino flux are presented.

\subsection{All sky scan}

\Autoref{fig:skymap} depicts the pre-trial p-value $\pval$ of all points in the
sky in equatorial coordinates (J2000) with respect to the null-hypothesis of no
observed clustering.

In the northern sky, the most significant position was at $\alpha=32.2\deg$,
$\delta=62.1\deg$ at an accuracy of $0.35\deg$ ($0.5\deg$) for $1\sigma$
($90\%$) contours using Wilks' theorem with two degrees of freedom. The best fit
parameters at the location are $\hatns=32.6$ and $\hat{\gamma}=2.8$, yielding a
pre-trial p-value of $1.82\times10^{-6}$.  Looking at each of the combined
seasons individually reveals that for each season clustering is observed,
providing no indication of time-dependence that could suggest additional
evidence for an astrophysical origin.

\begin{figure}[t]
    \centering
    \includegraphics[width=\columnwidth]{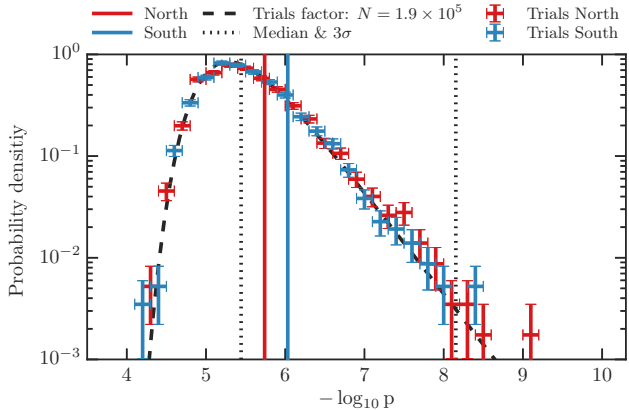}
    \caption{Trial correction of the most significant spots in the sky that
             were observed in the seven year search. Solid vertical lines
             indicate the pre-trial p-value of the most significant spots in
             each half of the sky; crosses show the distribution of spots
             similarly obtained in scrambled data trials. The trials are
             modeled by an analytic parameterization of the trial correction
             (\Autoref{eq:trial_corr}, black dashed line) that corresponds to
             $1.9\times10^5$ independent trials per half of the sky.}
    \label{fig:post-trial}
\end{figure}

In the southern sky, the most significant point is at $\alpha=174.6\deg$,
$\delta=-39.3\deg$. The best fit point is at $\hatns=15.4$, with spectral
index $\hat{\gamma}=2.9$. The uncertainty of the location amounts to
$0.22\deg$ ($0.32\deg$) for $1\sigma$ ($90\%$). The pre-trial p-value is
$0.93\times10^{-6}$; most of the significance at this location is shared by the
newly added data of through-going and starting tracks. Indeed, one starting
track is within $0.9\deg$ distance to the location which is wihtin $1\sigma$ of
its reconstruction uncertainty.

\begin{figure}[t]
    \centering
    \includegraphics[width=\columnwidth]{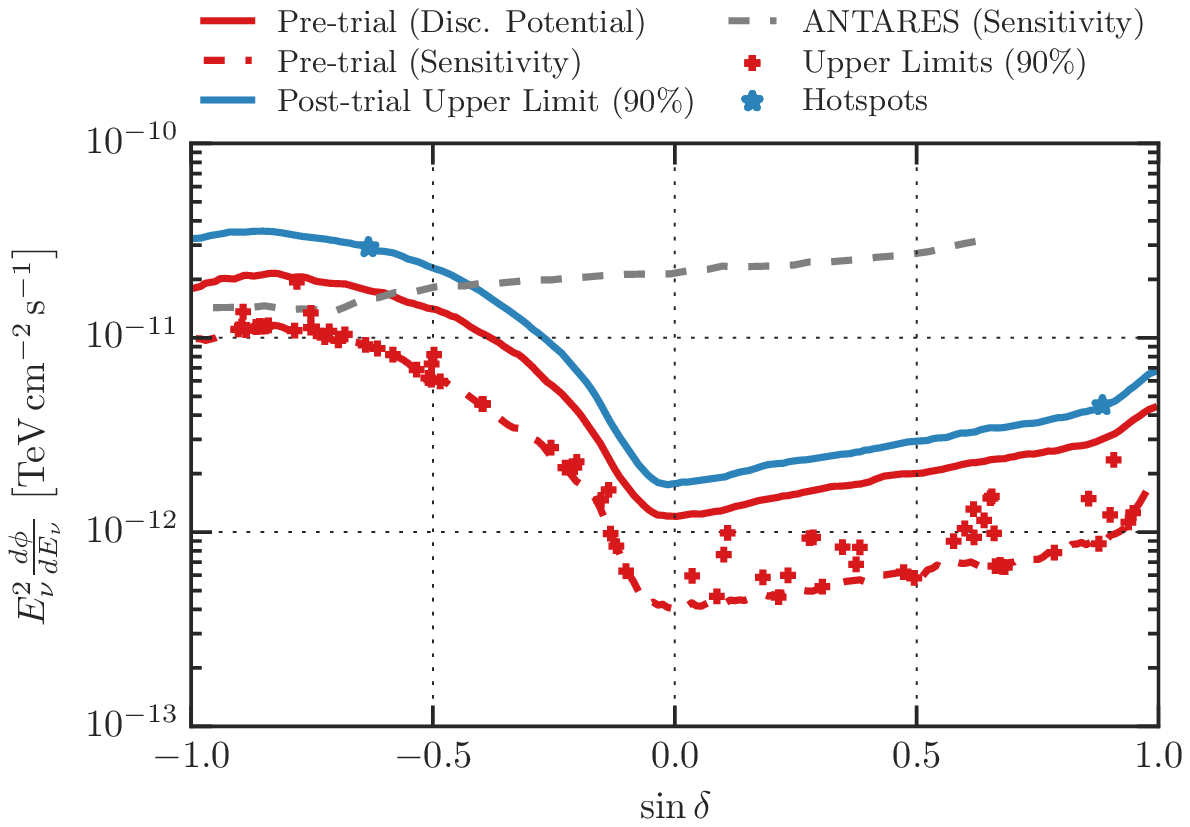}
    \caption{Discovery potential ($5\sigma$, solid red) and sensitivity (dashed
             red) for a $\nu_\mu+\bar{\nu}_\mu$ unbroken $\Edpde$ flux shown
             against declination $\delta$. The gray line shows the results
             of~\citep{Adrian-Martinez:2014wzf} in the south. Upper limits of
             source candidates in \autoref{tab:sourcelist1} and
             \autoref{tab:sourcelist2} are depicted by red crosses. The blue
             line represents the upper limit for the observed most significant
             spots in each half of the sky for all declinations, the actual
             declination position of the spots is indicated by a star.}
    \label{fig:sensitivity}
\end{figure}

Due to the large number of tested locations in the sky, the two most significant
locations in the sky have to be trial corrected with the trial correction in
\autoref{eq:trial_corr} that is estimated by repeating the full-sky scan on
scrambled data trials, as shown in \autoref{fig:post-trial}. This yields
post-trial p-values of $29\%$, $17\%$ for northern and southern sky,
respectively. Hence, the full-sky results are in agreement with a pure
background assumption, and no significant clustering is observed. For an
unbroken $E^{-2}$ power-law spectrum, the \UL s of the two most significant
positions are $\Edpde=4.49\times10^{-12}\:\flux$ in the northern sky, and
$\Edpde=2.92\times10^{-11}\:\flux$ in the southern sky. For softer spectra of
$E^{-3}$, the \UL s yield $E^3\dpde=5.08\times10^{-11}\:\sflux$ and
$E^3\dpde=1.29\times10^{-8}\:\sflux$ for the northern and southern spot,
respectively.  In \autoref{fig:sensitivity}, the solid blue line indicates the
\UL\ established by the hottest spot results. A neutrino source at any
declination $\delta$ that would emit a steady flux higher than this curve,
would be detected $90\%$ of the time as having a greater significance than that
actually observed for the hottest spots found in the analysis (whose \UL s are
highlighted as stars on the blue line).

\begin{figure*}[t]
    \centering
    \includegraphics[width=\columnwidth]{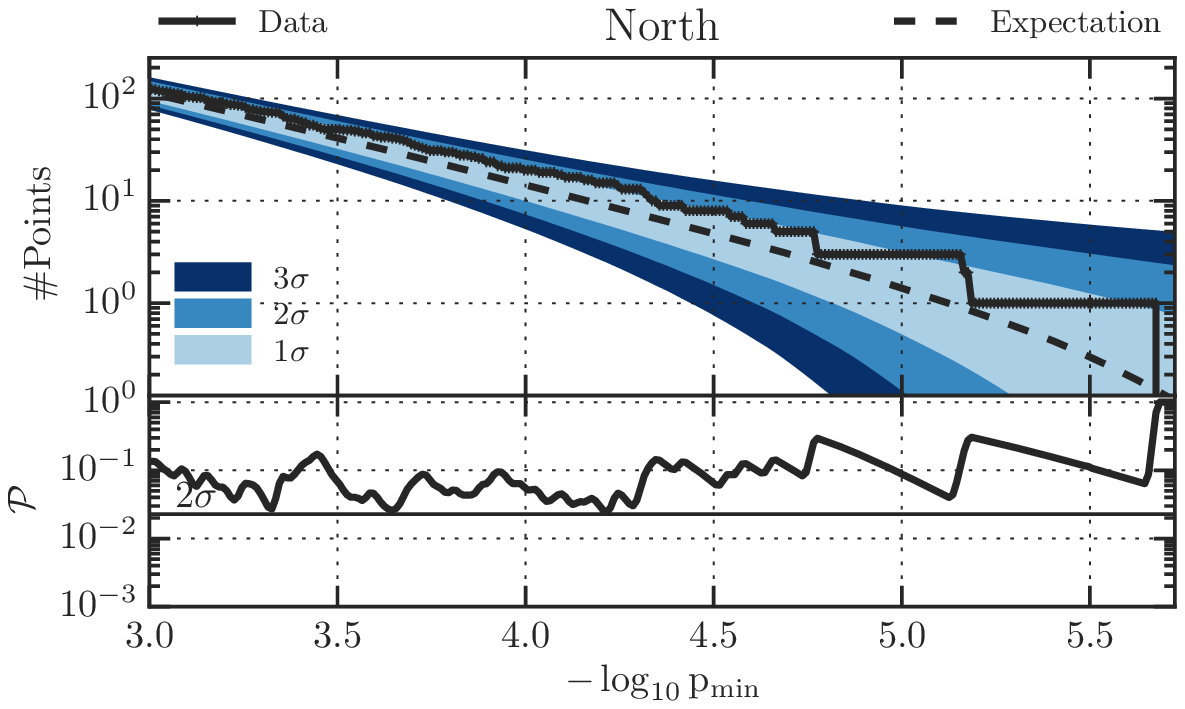}
    \includegraphics[width=\columnwidth]{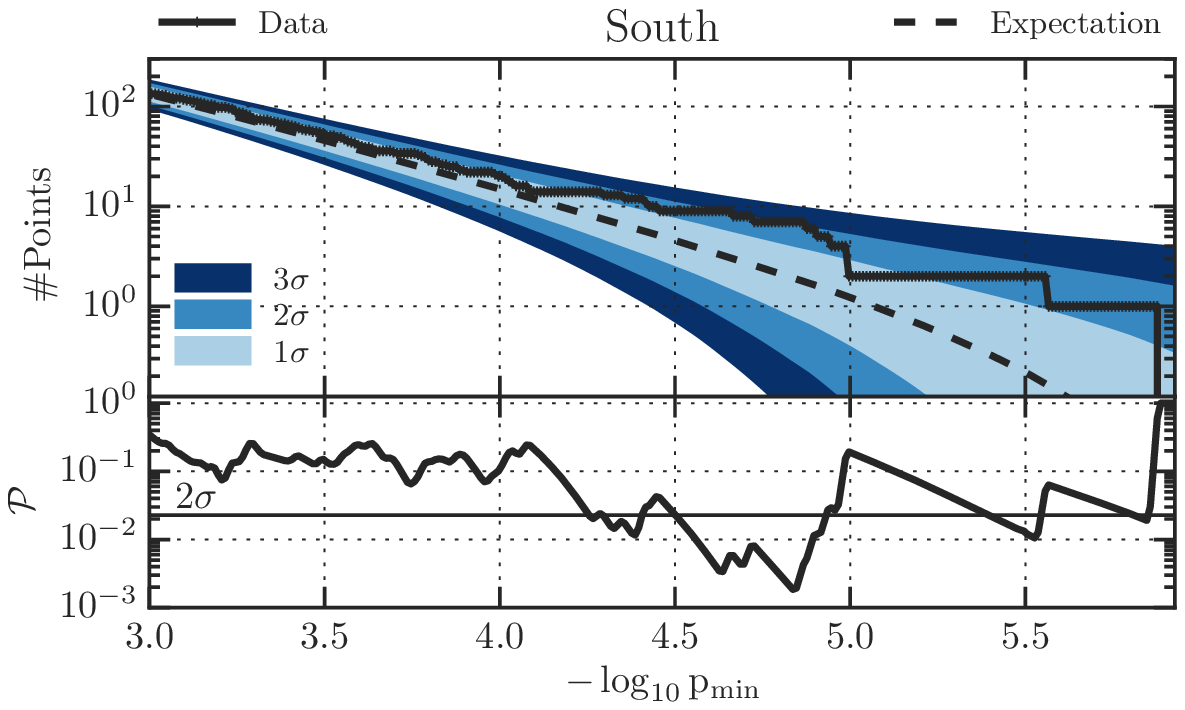}
    \caption{Number of associated spots observed with a minimum pre-trial
             p-value of $\pval_\mathrm{min}$ of the unbinned likelihood fit in
             the northern (left) and southern sky (right), respectively. The
             top plots show the number of spots (solid black) in comparison to
             background expectation (dashed black) with confidence intervals of
             one, two, and three standard deviations shown as shaded blue
             areas.  The points on the far right end of the x-axis correspond
             to the hottest spots observed in the northern and southern sky.
             The bottom plots show the local p-value $\mathcal{P}$
             (\Autoref{eq:HPApval}) of observing \emph{X} events given the
             background mean expectation of \emph{Y} using Poissonian
             statistics.}
    \label{fig:population}
\end{figure*}

Besides the results of the full-sky scan, there are two neutrino events
detected with IceCube that are worth commenting on here. The first one is
the highest energetic neutrino event detected ($4.5\pm1.2\:\mathrm{PeV}$) so
far with IceCube~\citep{Schoenen:2015abc, Aartsen:2016xlq}, a neutrino-induced
up-going muon track with very precise angular resolution. This neutrino event
is part of the through-going track sample (\Autoref{sec:thru}). At its position
($\alpha=110\deg$, $\delta=11.5\deg$), no significant clustering is observed
(pre-trial $5.2\%$). A slight excess is indeed observed, but originates from
the PeV event alone.  The second interesting event is a straight down-going
starting track at $430\:\mathrm{TeV}$ deposited energy~\citep{Aartsen:2015zva}.
Not only does it start inside of the IceCube detector, but the reconstructed
track points back to the IceTop surface detector and no atmospheric shower is
observed in coincidence with the event. This event is part of the starting
track sample (\Autoref{sec:starting}), but no clustering of events apart from
the track itself is observed at the location in the sky ($\alpha=218\deg$,
$\delta=-86\deg$) and the pre-trial p-value is~$0.6\%$.

\subsection{Hotspot population}

\begin{figure*}[t]
    \centering
    \includegraphics[width=\columnwidth]{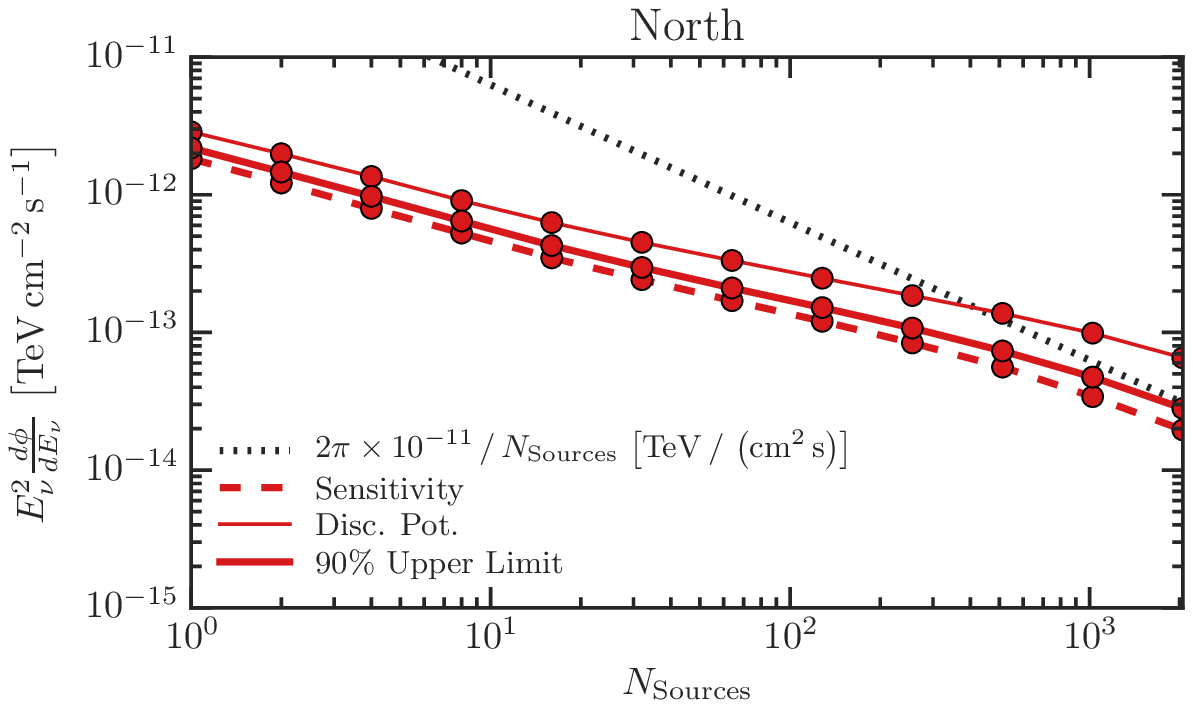}
    \includegraphics[width=\columnwidth]{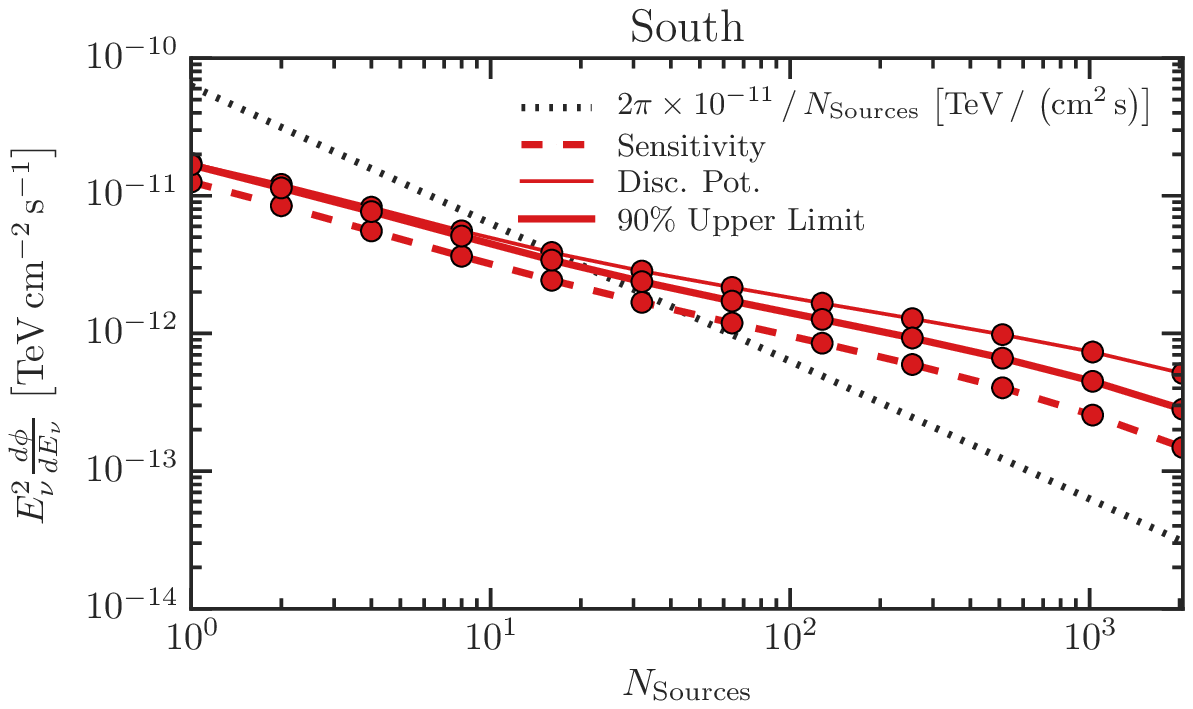}
    \caption{Discovery potential ($5\sigma$, solid thin line), sensitivity
             (dashed line), and \UL\ (solid thick line) for an
             unbroken $E^{-2}$ spectrum in the population analysis for northern
             (left) and southern (right) sky, respectively. The flux is shown
             per source for increasing number of sources distributed uniformly
             on half of the sky. The black dashed line shows the
             IceCube-measured astrophysical neutrino flux per source if it was
             distributed homogeneously among $N_\mathrm{Sources}$ sources in
             this half of the sky.}
    \label{fig:pop-sens}
\end{figure*}

The search for populations of weak sources in the full-sky in
\autoref{fig:skymap} did not reveal any significant outcome above background
expectation. In \autoref{fig:population}, the number of spots versus pre-trial
p-value $\pval_\mathrm{min}$ threshold is shown for northern (left) and
southern sky (right). The observed number of spots is shown versus background
expectation with shaded areas indicating $1\sigma$, $2\sigma$, and $3\sigma$
intervals. This is then converted to a \emph{local} p-value $\mathcal{P}$
according to \autoref{eq:HPApval}.

In the northern sky, the most significant excess is observed above a threshold
of $\pval_\mathrm{min}\geq3.35$ with $72$ spots above a scrambled data
expectation of $56.7$. The local p-value of such an excess is
$\mathcal{P}=2.8\%$ and increases to $25\%$ after trial correction. For the
southern sky, $7$ spots above an expectation of $2.1$ spots at
$\pval_\mathrm{min}\geq4.66$ are reported. The probability of this happening by
fluctuations of background is $0.62\%$. After correcting this for trials by
scanning in $\pval_\mathrm{min}$ for the biggest excess, the p-value increases
to $8.2\%$.  Restricting the analysis only to regions within $\pm15\deg$ of the
Galactic Plane the biggest excess is observed at $\pval_\mathrm{min}\geq5.68$,
corresponding to a single spot, which is also the same spot that was most
significant in the northern hemisphere full sky scan.  The background
expectation is $0.04$ giving a local p-value of $3.8\%$ that increases to
$26\%$ after trial correction.

Consequently, in both hemispheres and the Galactic Plane, no significant
population of sources over background expectation was found.
\autoref{fig:pop-sens} shows the sensitivity, $5\sigma$ discovery potential,
and \UL\ for northern (left) and southern sky (right). This is shown against an
increasing number of sources, where each source is assumed to be of equal
luminosity with an unbroken $E^{-2}$ spectrum. The flux value is averaged over
all declinations of the corresponding half of the sky. By comparing this to a
scenario where all of the observed diffuse astrophysical neutrino
flux~\citep{Aartsen:2014gkd} $\Edpde\sim10^{-11}\flux\,\mathrm{sr^{-1}}$ is
shared equally among $N_\mathrm{Sources}$ in the sky, in the northern sky, the
analysis result excludes populations of 1000 or fewer equal-strength sources,
whereas in the south, the result only excludes populations of 40 or fewer
equal-strength sources of the astrophysical flux.

\subsection{Source list candidates}

In \autoref{tab:sourcelist1} and \autoref{tab:sourcelist2}, the fit results of
the two source lists are quoted. For each source, the fit parameters $\hatns$
and $\hat{\gamma}$ are quoted. Furthermore, the pre-trial p-value and the mean
number of background events $B_{1\deg}$ within a one-degree circle around the
source is listed. For each source, \UL s are calculated using an unbroken
$E^{-2}$ spectrum. The upper-limits are shown as red crosses in
\autoref{fig:sensitivity} at the corresponding declination of the source.

In the first source list, the sources most significant are the blazar
\OES\ and flat spectrum radio quasar \PKS\ in the northern,
southern sky, respectively. At the position of \OES\ the pre-trial p-value is
$1.8\%$ with best fit-parameters $\hatns=15.4$ and $\hat{\gamma}=3.1$. The
resulting \UL\ for an unbroken $E^{-2}$ $\nu_\mu+\bar{\nu}_\mu$ flux of the
observed p-value at declination $\delta=65.15\deg$ of \OES\ yields
$\Edpde=2.36\times10^{-12}\flux$. For \PKS\ at $\delta=-7.87\deg$, the
pre-trial p-value of $5.3\%$ with $\hatns=7.3$ and $\hat{\gamma}=2.6$ results
in an \UL\ of $\Edpde=1.65\times10^{-12}\flux$. For trial correction, the
source list is split in a northern and southern part, with the division at
$\delta=-5^\circ$. The size of the source lists is then 34 (10) and yields a
trial corrected p-value of $54\%$ ($37\%$) for the northern sky (southern
sky). Hence, the results of the first source list are in agreement with
background expectation.

In the second source list, the most significant source is \HESS. The fit values
$\hatns=2.4$ and $\hat{\gamma}=4.0$ result in a pre-trial p-value of
$0.22\%$. The \UL\ is $\Edpde=1.94\times10^{-11}\flux$. Trial correction with
all 30 sources of the list yields a post-trial p-value of $9.3\%$. Most of the
significance at the position of \HESS\ comes from one starting track
only $0.34\deg$ away, while no significant clustering of high-energy events
is observed in the through-going event samples. Starting tracks access lower
energies in the southern sky (cf. \autoref{fig:effA}, \autoref{fig:diff-sens}).
As explained in \autoref{eq:nsj}, the number of source-like neutrinos $\ns$
is distributed among the different samples according to their signal
expectation for a spectral index $\gamma$.
Consequently, if the clustering is only observed in starting tracks, soft
spectral indices at the boundary $\hat{\gamma}=4$ are preferred as they give
more weight to the starting track sample that is most efficient for soft
spectra compared to through-going track samples in the southern sky.

In conclusion, both of the two source lists show no significant evidence for
clustering of astrophysical neutrinos, and all results are consistent with
background.

\begin{figure}[t]
    \centering
    \includegraphics[width=\columnwidth]{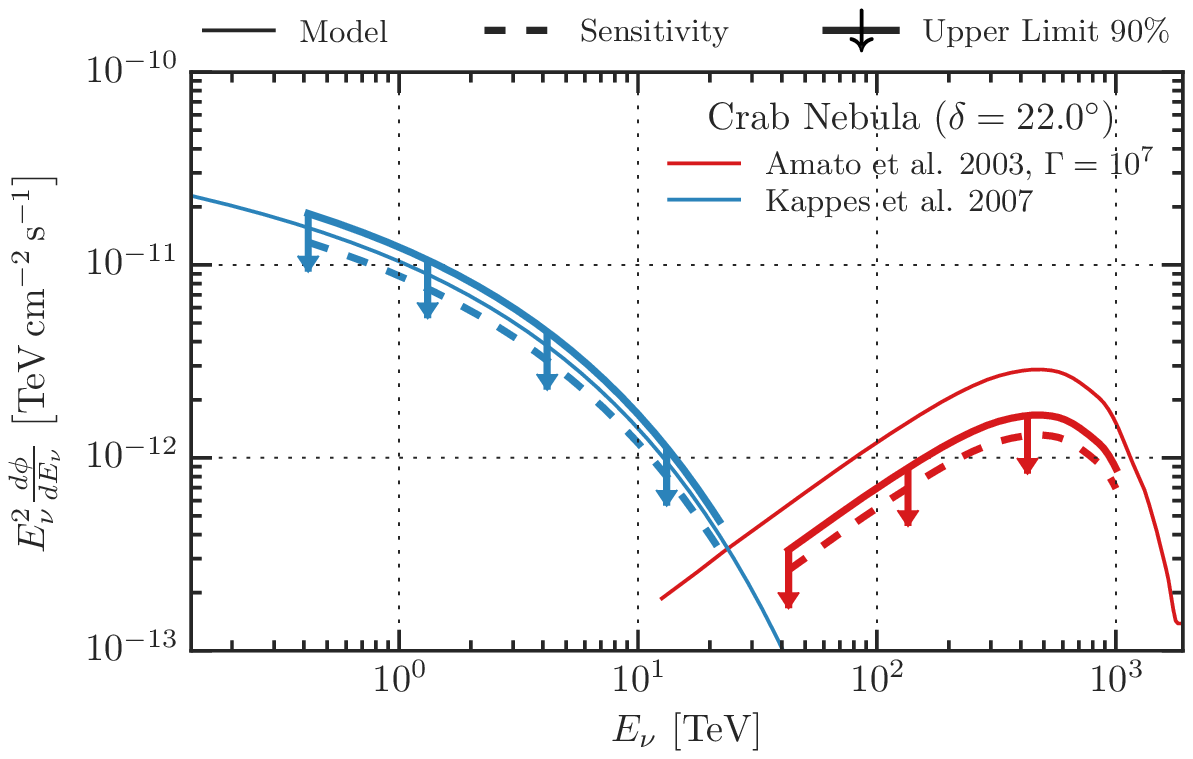}
    \caption{Differential $\nu_\mu+\bar{\nu}_\mu$ energy spectra versus
             neutrino energy for the Crab nebula. The figure shows the
             conversion of the observed gamma-ray flux of the Crab nebula to
             neutrinos~\citep{Kappes:2006fg} (blue), and a simulation of inelastic
             p-p scattering at the source~\citep{Amato:2003kw} (red). Thick lines
             correspond to the \UL\ of this search, thin lines
             are represent the model. The sensitivity of this analysis is shown
             as dashed line. \UL\ and sensitivity are shown for the energy
             interval, where $90\%$ of the events originate that are most
             signal-like, cf. \autoref{fig:diff-sens}.}
    \label{fig:crab}
\end{figure}

\subsection{Multi-wavelength model constraints}

Above, upper limits on neutrino emission from sources were made using unbroken
$\dpde\propto E^{-2}$ fluxes as benchmark.  However, more specific estimates
for neutrino fluxes can be made using multi-wavelength data of $\gamma$-ray
sources. In decays of pions, both $\gamma$-rays as well as neutrinos are
produced~\citep{Anchordoqui:2013dnh}. Due to long-baseline oscillations, any
flavor composition at the source will result in a sizable fraction of muon
neutrinos at Earth~\citep{Athar:2005wg}. Due to no significant observation of
clustering, upper-limits on specific models are set by injecting signal events
at the corresponding source declination according to the energy spectrum
$\Edpde$ given by the model.

The first source considered is the Crab Nebula, a pulsar wind nebula,
and the strongest steady TeV $\gamma$-ray source in the sky. At a declination
of $\delta=22\deg$, it is in the region where IceCube covers a wide range of
energies efficiently, compare \autoref{fig:diff-sens}. Two scenarios of
neutrino emission from this source are considered. \Autoref{fig:crab}
shows the neutrino emission (thin line) with respect to the \UL\ (thick line)
of IceCube. At the position of the Crab Nebula, an over-fluctuation with
p-value $34\%$ is observed which results in an upper-limit higher than
IceCube's potential sensitivity (dashed).  By convoluting the differential
discovery potential (\Autoref{fig:diff-sens}) at the source position and the
model neutrino spectrum, the energy-region, where $90\%$ of the constraining
power of IceCube originates, for a specific model is calculated. This is
indicated by the energy region where the limits and sensitivities are drawn in
\autoref{fig:crab} for each of the following models.

\begin{figure}[t]
    \centering
    \includegraphics[width=\columnwidth]{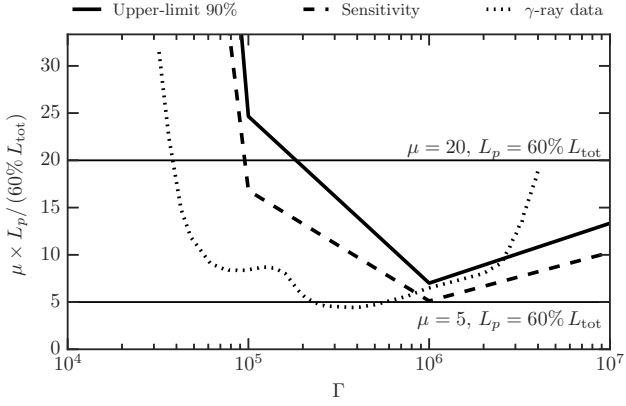}
    \caption{Sensitivity and \UL\ (dashed and solid lines, respectively) for
             the proton wind luminosity $L_p$ and target
             density in \autoref{eq:target_density} of the Crab Nebula in
             \cite{Amato:2003kw} for different Lorentz factors $\Gamma$. Values
             of $\mu=20$ and $\mu=5$ ($L_p=60\%\,L_\mathrm{tot}$) are indicated
             as horizontal lines that correspond to upper limits of the target
             density and a uniform mass distribution in the nebula,
             respectively. The dotted line indicates
             upper-limits from gamma-ray observations compared to
             $\pi^0\to\gamma\gamma$ decays~\citep{Amato:2003kw,
             deJager:1996abc, Aharonian:2000pz}.}
    \label{fig:amato_gamma}
\end{figure}

Regarding the Crab Nebula, the first model taken into account (red in
\autoref{fig:crab}) is by \cite{Amato:2003kw} and uses inelastic p-p
scattering at the source to model the neutrino emission where
$60\%$ of the wind luminosity $L_\mathrm{tot}$ is carried by protons. The
model shown is for a Lorentz factor of the wind of $\Gamma=10^7$ where the
energy density peaks at the $\sim500\:\mathrm{TeV}$ and assumes a target
density of
\begin{align}
    n_t=10\mu\left(M_{N\odot}/R^3_\mathrm{pc}\right)\:\mathrm{cm^{-3}}
    \label{eq:target_density}
\end{align}
with $\mu=20$ (shown in \autoref{fig:crab}) as defined in Eq.~$9$ in
\cite{Amato:2003kw}. $M_{N\odot}$ and $R_\mathrm{pc}$ are the
mass of the supernova ejecta contained in filamentary
structures within the nebula and its radius, respectively. With the increased
statistics compared to the previous analysis~\citep{Aartsen:2014cva}, the \UL\
now surpasses the prediction of $\mu=20$, hence, constraining the target
density $\mu^{90\%}<12$.  Upper limits on $\mu$ and the proton wind luminosity
$L_p$ in units of the total wind luminosity $L_\mathrm{tot}$ for lower values
of the Lorentz wind $\Gamma$ are shown in \autoref{fig:amato_gamma}. Lower
values of $\Gamma$ shift the neutrino energy spectrum to lower energies into
the TeV region where IceCube is most sensitive in the Northern hemisphere.
Hence,  target densities $\mu<20$ as mentioned in \cite{Atoyan:1996abc} are
partially constrained by IceCube for a proton luminosity
fraction of $60\%$.  For $\Gamma=10^5$, the sensitivity is at the level of
$\mu=5$ corresponding to a uniform mass distribution in the nebula as pointed
out by \cite{Amato:2003kw}.
For Lorentz wind factors $\Gamma<10^6$, the strongest
constraint on the proton luminosity is given by gamma-ray observations via
neutral pion decay $\pi^0\to\gamma\gamma$~\citep{Amato:2003kw, deJager:1996abc,
Aharonian:2000pz}. For $\Gamma\geq10^6$, neutrino observations give additional
information and large values of $\Gamma=10^7$ are exclusively observable by
neutrinos due to the high energy spectrum.  \cite{DiPalma:2016yfy} revise the
predictions for PWNe given new TeV $\gamma$-ray
observations~\citep{Aliu:2008hc, Aliu:2012uj}, predicting $\sim13.09$ neutrino
events from the Crab nebula per year in the energy range from 1 to 100~TeV.
IceCube's \UL\ at the position of the Crab nebula (see
\autoref{tab:sourcelist1}) constrains the number of neutrinos from the source
to not more than 7.0 or 21.2 for an unbroken spectrum of $E^{-2}$ and $E^{-3}$
respectively from a total of seven years of data-taking. This is much lower
than the predicted number of neutrino events during the same time range.

The second neutrino spectrum tested for the Crab Nebula is the one
by~\citep{Kappes:2006fg} (blue). The neutrino spectrum is connected to
$\gamma$-rays assuming that both spectra originate from the same
pion-component. Thus, the resulting neutrino spectrum is fitted to an
exponentially cut-off power-law $\dpde\propto
E^{-2.15}\exp\left(-\sqrt{E_\nu/1.72\:\mathrm{TeV}}\right)$. Due to the
observed over-fluctuation at the position of the Crab Nebula, the \UL\
regarding this model exceeds the benchmark model of $100\%$ pion contribution
by a factor of 1.18 and is thus not constraining the amount of hadronic
acceleration. Nevertheless, IceCube's sensitivity (blue dashed line in
\autoref{fig:crab}) is $15\%$ below the model. Hence, with seven years
exposure, IceCube is now also sensitive to neutrino fluxes comparable to that
of the Crab nebula in gamma-rays, i.e. bright but falling off at relatively
low energies of a few TeV.

\begin{figure}[t]
    \centering
    \includegraphics[width=\columnwidth]{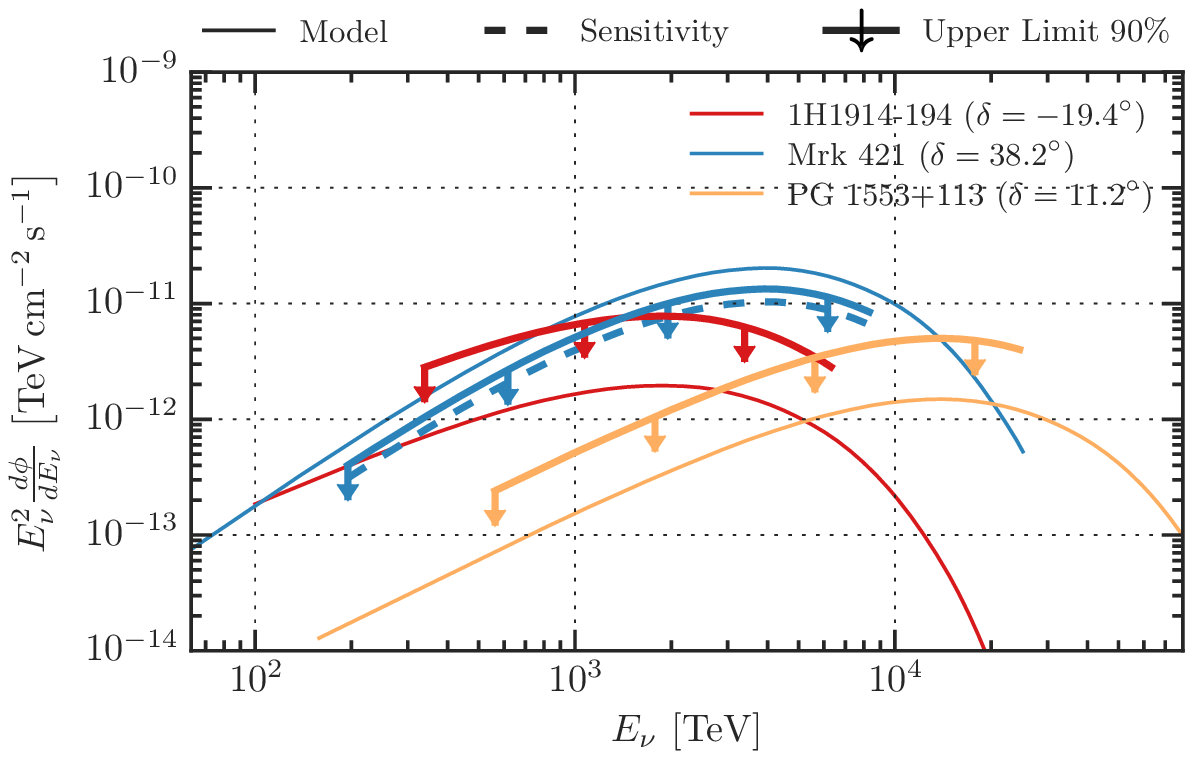}
    \caption{Same as \autoref{fig:crab} for blazars of type BL Lac modeled in
             \cite{Petropoulou:2015upa}.}
    \label{fig:mrk421}
\end{figure}

Another promising category of neutrino emitters are active galactic nuclei
(AGN), especially blazars of type BL Lac, as for example highlighted
in~\citep{Padovani:2014bha}. In \cite{Glusenkamp:2015jca}, an analysis of
IceCube neutrino events searched for correlations with blazars detected by
Fermi-LAT to probe their possible connection to the observed diffuse
astrophysical neutrino signal.  It was found that such AGN cannot be the
dominant contribution to the diffuse neutrino flux over the entire probed
energy range, but BL Lacs could possibly explain the high-energy part of the
flux~\citep{Padovani:2015mba}.  In \cite{Petropoulou:2015upa}, BL Lacs that
were found to be in spatial and energetic agreement with IceCube high-energy
starting events were modeled using lepto-hadronic fitting of multi-wavelength
data. Of the six modeled blazars, three are far from being
constrained by IceCube due to neutrino absorption in the Earth (1ES~1011+496)
or large atmospheric background deep in the southern sky (H2356-309,
1RXS~J05435-5532). \Autoref{fig:mrk421} shows the expected neutrino energy
spectra for the three remaining BL Lacs. The \UL\ for
1H1914-194, also in the southern sky, is a factor 3.9 above the prediction. For
PG1553+113, the contribution of proton-pion interactions in the model is small
(total neutrino flux relative to TeV $\gamma$-ray flux $Y_{\nu\gamma}=0.1$,
Eq.~13 in \cite{Petropoulou:2015upa}) resulting in a lower neutrino luminosity
and a \UL\ of a factor of three above the prediction.

The third model source considered here is Mrk~421 (blue) and shows a different
picture. Mrk~421 is one of the closest blazars with redshift $z\sim0.03$ and
with $\delta=38.2\deg$, it is located in the up-going region where IceCube is
most sensitive. Moreover, in \cite{Petropoulou:2015upa}, the multi-wavelength
data observed can be explained with a very high hadronic component
($Y_{\nu\gamma}=0.7$), thus, realizing a high neutrino luminosity. For such a
model, the \UL\ obtained by IceCube is $2/3$ of the predicted neutrino flux.
Hence, using the results at the position of Mrk~421, the hadronic acceleration
cannot be bigger than $Y_{\nu\gamma}\leq0.47$ assuming a steady emission over
the seven years of analyzed data, and thus constrains proton luminosity of the
source for steady emission or emission taking into account variability of the
source~\citep{Petropoulou:2015upa, Petropoulou:2016ujj}.

\begin{figure}[t]
    \centering
    \includegraphics[width=\columnwidth]{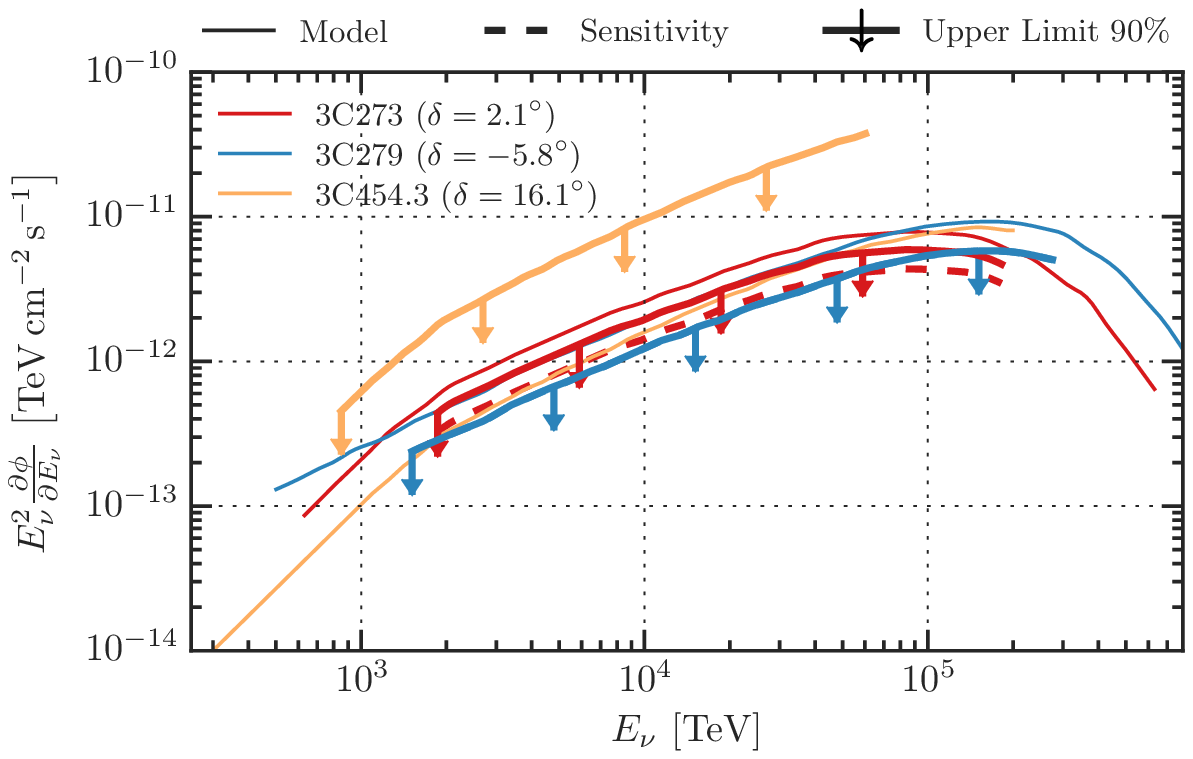}
    \caption{Same as \autoref{fig:crab} for blazars modeled in
             \cite{Reimer:2015abc}.}
    \label{fig:reimer_a}
\end{figure}

Other models for active galactic nuclei of type flat spectrum radio quasar or
BL Lac were modeled in \cite{Reimer:2015abc}. In \autoref{fig:reimer_a}, \UL s for
3C273 (red), 3C279 (blue), and 3C454.3 (yellow) are shown. The models in this
reference are very characteristic due to their hard spectra extending beyond
$10\:\mathrm{PeV}$ before cutting off. Hence, IceCube cannot constrain the
prediction for FSRQ 3C454.3 ($\delta=16.1\deg$) because most of the flux is
absorbed in the Earth and the \UL\ is a factor of five higher than the
prediction.  The two other FSRQ on the other hand are close to the horizon
where no absorption is present.  Therefore, the \UL\ placed by IceCube is lower
than the prediction, that is, $13\%$ and $40\%$ for 3C273 and 3C279,
respectively. Similar to Mrk~421 in the previous paragraph, assuming that the
emission is constant over the livetime of IceCube, this results constrains the
amount of hadronic acceleration possible due to a non-observation of neutrinos
by IceCube with respect to the model prediction. The other AGN
modeled in \cite{Reimer:2015abc} are in the northern sky and due to absorption,
the \UL\ by IceCube is at least a factor of 60 above the predicition.

Other models that were tested include galactic source like supernovae
remnants~\citep{Mandelartz:2013ht} and unidentified TeV sources in the
galaxy~\citep{Fox:2013oza} that are mainly in the southern sky
which coincides with most of the Galactic Plane. There, IceCube has to cope
with large atmospheric muon backgrounds and has reduced sensitivity with high
energy-thresholds, compare \autoref{fig:diff-sens} and
\autoref{fig:sensitivity}. Consequently, the current \UL s are at least a
factor of five (G40.5-05, \cite{Mandelartz:2013ht}) up to more than a hundred
(Vela X, \cite{Kappes:2006fg}) above the prediction. At these energies in the
southern sky region, neutrino telescopes located in the northern hemisphere can
place stronger constraints~\citep{Adrian-Martinez:2015ver, Trovato:2016oso}.

\section{Conclusions}
\label{sec:conclusion}

\begin{figure}[t]
    \centering
    \includegraphics[width=\columnwidth]{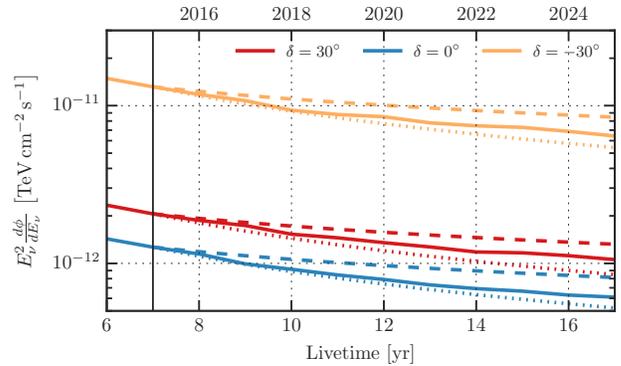}
    \caption{Time evolution in discovery potential ($5\sigma$) with increasing
             detector exposure. Different declinations are shown in different
             colors. The current status of this analysis corresponds to seven
             years or the year 2015 (indicated by vertical black line). Dashed
             lines indicate the sensitivity growth assuming a dependence
             with~$1/\sqrt{T}$, dotted with~$1/T$.}
    \label{fig:time-evo}
\end{figure}

Analyzing the full sky for clustering of astrophysical neutrinos using seven
years of through-going muon data from 2008 to 2015 and five years of starting
tracks (2010-2015), we did not find any significant steady point-like emission
over background expectation.

An unbiased scan of the full-sky was performed, as well as dedicated searches
using promising $\gamma$-ray candidates and searches for populations of weak
sources. Results are consistent with background and \UL s on steady neutrino
emission are calculated from the observations. In the northern sky
($\delta\geq5\deg$), the best limits are set by IceCube and the sensitivity
is below $\Edpde<10^{-12}\flux$ over a wide declination for the first time, compare
\autoref{fig:sensitivity}. Moreover, the declination dependent flux limit is a
factor of $\sim100$ ($\sim10$) below the integrated diffuse astrophysical
neutrino flux in the northern (southern) sky. With increased statistics of
three additional years compared to the previous analysis, the upper-limits in
the southern sky are of the same level as for \cite{Adrian-Martinez:2014wzf},
however, testing a complementary energy region above PeV neutrino energies.
Model-specific \UL s were calculated for representative cases from the
literature, and previously unconstrained neutrino emission scenarios for
sources in the northern sky are now disfavored by IceCube.
With the newly available data presented here, improved results
for other tests are anticipated, for example searches for extended sources or
stacking and time-dependent analyses~\citep{Aartsen:2014cva, Aartsen:2015wto}.

With increasing exposure, IceCube continues to improve the sensitivity to
steady neutrino fluxes. \Autoref{fig:time-evo} shows the progression of the
$5\sigma$ discovery potential with increasing time for three declinations. A
continuous gain in performance is observed with time $T$ that is faster than
$1/\sqrt{T}$, a scenario assuming statistical limitation by background, and
much closer to $1/T$ indicating limitation by signal statistics. This can
increase even further with anticipated improvements in background rejection,
angular reconstruction, or detector improvements~\citep{Aartsen:2014njl}.

\acknowledgments

We acknowledge the support from the following agencies:
U.S. National Science Foundation-Office of Polar Programs,
U.S. National Science Foundation-Physics Division,
University of Wisconsin Alumni Research Foundation,
the Grid Laboratory Of Wisconsin (GLOW) grid infrastructure at the University of Wisconsin - Madison, the Open Science Grid (OSG) grid infrastructure;
U.S. Department of Energy, and National Energy Research Scientific Computing Center,
the Louisiana Optical Network Initiative (LONI) grid computing resources;
Natural Sciences and Engineering Research Council of Canada,
WestGrid and Compute/Calcul Canada;
Swedish Research Council,
Swedish Polar Research Secretariat,
Swedish National Infrastructure for Computing (SNIC),
and Knut and Alice Wallenberg Foundation, Sweden;
German Ministry for Education and Research (BMBF),
Deutsche Forschungsgemeinschaft (DFG),
Helmholtz Alliance for Astroparticle Physics (HAP),
Research Department of Plasmas with Complex Interactions (Bochum), Germany;
Fund for Scientific Research (FNRS-FWO),
FWO Odysseus programme,
Flanders Institute to encourage scientific and technological research in industry (IWT),
Belgian Federal Science Policy Office (Belspo);
University of Oxford, United Kingdom;
Marsden Fund, New Zealand;
Australian Research Council;
Japan Society for Promotion of Science (JSPS);
the Swiss National Science Foundation (SNSF), Switzerland;
National Research Foundation of Korea (NRF);
Villum Fonden, Danish National Research Foundation (DNRF), Denmark

\bibliography{references}

\onecolumngrid
\newpage

\begin{deluxetable*}{llrrlrcrr}

\tabletypesize{\small}

\tablecaption{Sources contained in the first source list. In addition to its
              type, common name, and coordinates (Equatorial (J2000)), the
              best-fit of signal events $\ns$ and spectral index $\gamma$ with
              the pre-trial p-value $-\log_{10}\mathrm{p}$ are quoted. For null
              observations $\ns=0$, no p-value and spectral index are quoted. A
              \UL\ for an $E^{-2}$ unbroken power-law is calculated for all
              sources that showed clustering of neutrinos. The upper limits are
              shown in \autoref{fig:sensitivity}, for null-fits the limit
              equals the sensitivity of this analysis at the corresponding
              declination.\label{tab:sourcelist1}}

\tablehead{\colhead{Type}
           & \colhead{Source}
           & \colhead{$\alpha$}
           & \colhead{$\delta$}
           & \colhead{p-value}
           & \colhead{$\ns$}
           & \colhead{$\gamma$}
           & \colhead{$B_{1^\circ}$}
           & \colhead{$\Phi^{90\%}_{\nu_\mu+\bar{\nu}_\mu}$}
           \\
           \colhead{}
           & \colhead{}
           & \colhead{$1^\circ$}
           & \colhead{$1^\circ$}
           & \colhead{}
           & \colhead{}
           & \colhead{} & \colhead{}
           & \colhead{$\mathbf{\zeta}$\tablenotemark{a}}}

\startdata
BL Lac               & PKS 0537-441         &   84.71 &  -44.09 & \nodata &  0.0 & \nodata & 45.7 &  9.79\\
                     & PKS 2155-304         &  329.72 &  -30.23 & \nodata &  0.0 & \nodata & 52.6 &  6.07\\
                     & PKS 0235+164         &   39.66 &   16.62 & 0.12    & 16.2 & 3.4     & 72.0 &  0.94\\
                     & 1ES 0229+200         &   38.20 &   20.29 & 0.20    &  9.5 & 3.5     & 66.6 &  0.84\\
                     & W Comae              &  185.38 &   28.23 & \nodata &  0.0 & \nodata & 60.8 &  0.62\\
                     & Mrk 421              &  166.11 &   38.21 & 0.32    &  2.7 & 1.8     & 53.1 &  0.94\\
                     & Mrk 501              &  253.47 &   39.76 & 0.18    & 10.5 & 4.0     & 52.0 &  1.15\\
                     & BL Lac               &  330.68 &   42.28 & \nodata &  0.0 & \nodata & 50.4 &  0.63\\
                     & H1426+428            &  217.14 &   42.67 & \nodata &  0.0 & \nodata & 50.9 &  0.70\\
                     & 3C 66A               &   35.67 &   43.04 & \nodata &  0.0 & \nodata & 50.9 &  0.70\\
                     & 1ES 2344+514         &  356.77 &   51.70 & \nodata &  0.0 & \nodata & 46.3 &  0.81\\
                     & 1ES 1959+650         &  300.00 &   65.15 & 0.018\tablenotemark{b}   & 15.4 & 3.1     & 42.8 &  2.36\\
                     & S5 0716+71           &  110.47 &   71.34 & \nodata &  0.0 & \nodata & 38.4 &  1.34\\
Flat-spectrum        & PKS 1454-354         &  224.36 &  -35.65 & \nodata &  0.0 & \nodata & 49.1 &  7.99\\*
radio quasar         & PKS 1622-297         &  246.53 &  -29.86 & 0.11    &  3.8 & 2.3     & 52.7 &  8.20\\
                     & QSO 1730-130         &  263.26 &  -13.08 & \nodata &  0.0 & \nodata & 49.8 &  2.18\\
                     & PKS 1406-076         &  212.24 &   -7.87 & 0.053\tablenotemark{c}   &  7.3 & 2.6     & 50.5 &  1.65\\
                     & QSO 2022-077         &  306.42 &   -7.64 & \nodata &  0.0 & \nodata & 50.5 &  0.99\\
                     & 3C 279               &  194.05 &   -5.79 & 0.42    &  0.5 & 2.0     & 54.3 &  0.63\\
                     & 3C 273               &  187.28 &    2.05 & 0.25    &  7.7 & 3.2     & 76.4 &  0.59\\
                     & PKS 1502+106         &  226.10 &   10.49 & 0.38    &  3.1 & 2.7     & 73.7 &  0.59\\
                     & PKS 0528+134         &   82.73 &   13.53 & 0.44    &  2.7 & 4.0     & 73.0 &  0.60\\
                     & 3C 454.3             &  343.49 &   16.15 & 0.12    &  4.1 & 2.0     & 72.3 &  0.93\\
                     & 4C 38.41             &  248.81 &   38.13 & 0.12    &  6.3 & 2.4     & 53.2 &  1.31\\
Galactic center      & Sgr A*               &  266.42 &  -29.01 & \nodata &  0.0 & \nodata & 52.2 &  6.08\\
Not identified       & MGRO J1908+06        &  286.98 &    6.27 & 0.025   &  4.5 & 2.0     & 74.9 &  0.99\\
Pulsar wind          & Geminga              &   98.48 &   17.77 & \nodata &  0.0 & \nodata & 69.3 &  0.49\\*
nebula               & Crab Nebula          &   83.63 &   22.01 & 0.34    &  6.1 & 3.8     & 67.0 &  0.68\\
                     & MGRO J2019+37        &  305.22 &   36.83 & 0.23    &  7.8 & 4.0     & 53.2 &  1.04\\
Star formation region& Cyg OB2              &  308.30 &   41.32 & 0.26    &  5.9 & 4.0     & 50.6 &  0.99\\
Supernova remnant    & IC443                &   94.21 &   22.50 & 0.22    &  8.1 & 4.0     & 66.0 &  0.83\\
                     & Cas A                &  350.81 &   58.81 & 0.14    &  8.1 & 4.0     & 44.5 &  1.49\\
                     & TYCHO                &    6.36 &   64.18 & 0.27    &  4.6 & 3.4     & 42.4 &  1.23\\
Starburst/radio      & Cen A                &  201.37 &  -43.02 & 0.21    &  0.5 & 1.2     & 46.2 & 10.41\\*
galaxy               & M87                  &  187.71 &   12.39 & \nodata &  0.0 & \nodata & 73.2 &  0.48\\
                     & 3C 123.0             &   69.27 &   29.67 & \nodata &  0.0 & \nodata & 59.5 &  0.57\\
                     & Cyg A                &  299.87 &   40.73 & 0.068   &  2.1 & 1.4     & 51.1 &  1.50\\
                     & NGC 1275             &   49.95 &   41.51 & \nodata &  0.0 & \nodata & 50.6 &  0.71\\
                     & M82                  &  148.97 &   69.68 & \nodata &  0.0 & \nodata & 39.7 &  1.09\\
HMXB/mqso            & SS 433               &  287.96 &    4.98 & 0.40    &  4.1 & 4.0     & 75.8 &  0.47\\
                     & HESS J0632+057       &   98.24 &    5.81 & 0.10    & 13.6 & 3.6     & 75.4 &  0.77\\
                     & Cyg X-1              &  299.59 &   35.20 & 0.31    &  4.5 & 4.0     & 54.4 &  0.90\\
                     & Cyg X-3              &  308.11 &   40.96 & 0.077   & 12.8 & 4.0     & 51.3 &  1.53\\
                     & LSI 303              &   40.13 &   61.23 & \nodata &  0.0 & \nodata & 43.8 &  0.79\\
\enddata

\tablecomments{Sources of the list are grouped by their classification and
ordered in ascending declination $\delta$. High-mass X-ray binaries and
micro-quasars are abbreviated by \emph{HMXB/mqso}.\\
Spectral indices quoted as $\gamma=4.0$ are at the boundary of the parameter
space for minimization. This can happen when there is an over-fluctuation of
low-energy events close to the source location. It was ensured that for all
events, the minimization converged successfully to the minimum within the
parameter-space.}
\tablenotetext{a}{Upper limits are in units of
$\mathbf{\zeta}=10^{-12}\flux$}
\tablenotetext{b}{Most significant source in northern sky in this table.
The trial-corrected p-value is $54\%$.}
\tablenotetext{c}{Most significant source in southern sky in this table.
The trial-corrected p-value is $37\%$.}

\end{deluxetable*}

\begin{deluxetable*}{llrrlrcrr}

\tabletypesize{\small}

\tablecaption{Sources contained in the second source
              list, using only sources in the southern sky and focusing on
              galactic objects.The information listed is the same as in
              \autoref{tab:sourcelist1}.\label{tab:sourcelist2}}

\tablehead{\colhead{Type}
           & \colhead{Source}
           & \colhead{$\alpha$}
           & \colhead{$\delta$}
           & \colhead{p-value}
           & \colhead{$\ns$}
           & \colhead{$\gamma$}
           & \colhead{$B_{1^\circ}$}
           & \colhead{$\Phi^{90\%}_{\nu_\mu+\bar{\nu}_\mu}$}
           \\
           \colhead{}
           & \colhead{}
           & \colhead{$1^\circ$}
           & \colhead{$1^\circ$}
           & \colhead{}
           & \colhead{}
           & \colhead{} & \colhead{}
           & \colhead{$\mathbf{\zeta}$\tablenotemark{a}}}

\startdata
BL Lac               & PKS 2005-489         &  302.37 &  -48.82 & 0.071   &  0.9 & 1.0     & 44.7 & 13.45\\
                     & PKS 0426-380         &   67.17 &  -37.93 & \nodata &  0.0 & \nodata & 47.2 &  8.93\\
                     & PKS 0548-322         &   87.67 &  -32.27 & \nodata &  0.0 & \nodata & 51.2 &  6.79\\
                     & H2356-309            &  359.78 &  -30.63 & \nodata &  0.0 & \nodata & 52.1 &  6.18\\
                     & 1ES 1101-232         &  165.91 &  -23.49 & \nodata &  0.0 & \nodata & 52.6 &  4.64\\
                     & 1ES 0347-121         &   57.35 &  -11.99 & 0.21    &  1.4 & 2.4     & 52.2 &  2.16\\
Flat-spectrum        & PKS 0454-234         &   74.27 &  -23.43 & \nodata &  0.0 & \nodata & 52.8 &  4.58\\*
radio quasar         & PKS 0727-11          &  112.58 &  -11.70 & 0.20    &  2.7 & 3.7     & 52.0 &  2.30\\
Not identified       & HESS J1507-622       &  226.72 &  -62.34 & \nodata &  0.0 & \nodata & 43.4 & 11.02\\
                     & HESS J1503-582       &  226.46 &  -58.74 & \nodata &  0.0 & \nodata & 45.5 & 11.79\\
                     & HESS J1741-302       &  265.25 &  -30.20 & 0.19    &  2.1 & 4.0     & 52.6 &  7.33\\
                     & HESS J1834-087       &  278.69 &   -8.76 & 0.21    &  1.2 & 3.7     & 49.5 &  1.47\\
Pulsar wind          & HESS J1356-645       &  209.00 &  -64.50 & \nodata &  0.0 & \nodata & 42.4 & 10.90\\*
nebula               & PSR B1259-63         &  197.55 &  -63.52 & 0.21    &  1.3 & 2.0     & 42.7 & 11.43\\
                     & HESS J1303-631       &  195.74 &  -63.20 & 0.076   &  4.5 & 2.3     & 42.3 & 13.61\\
                     & MSH 15-52            &  228.53 &  -59.16 & \nodata &  0.0 & \nodata & 44.9 & 11.28\\
                     & HESS J1023-575       &  155.83 &  -57.76 & \nodata &  0.0 & \nodata & 46.4 & 11.79\\
                     & HESS J1616-508       &  243.78 &  -51.40 & 0.0022\tablenotemark{b}  &  2.4 & 4.0     & 45.0 & 19.37\\
                     & HESS J1632-478       &  248.04 &  -47.82 & 0.16    &  1.5 & 2.7     & 44.7 & 10.79\\
                     & Vela X               &  128.75 &  -45.60 & 0.13    &  2.7 & 2.4     & 45.8 & 10.79\\
                     & HESS J1837-069       &  279.41 &   -6.95 & \nodata &  0.0 & \nodata & 48.1 &  0.89\\
Supernova remnant    & RCW 86               &  220.68 &  -62.48 & \nodata &  0.0 & \nodata & 43.1 & 11.02\\
                     & RX J0852.0-4622      &  133.00 &  -46.37 & \nodata &  0.0 & \nodata & 45.6 & 10.40\\
                     & RX J1713.7-3946      &  258.25 &  -39.75 & \nodata &  0.0 & \nodata & 45.5 &  9.22\\
                     & W28                  &  270.43 &  -23.34 & \nodata &  0.0 & \nodata & 52.9 &  4.58\\
Seyfert galaxy       & ESO 139-G12          &  264.41 &  -59.94 & \nodata &  0.0 & \nodata & 44.0 & 11.28\\
HMXB/mqso            & Cir X-1              &  230.17 &  -57.17 & \nodata &  0.0 & \nodata & 46.3 & 11.03\\
                     & GX 339-4             &  255.70 &  -48.79 & 0.15    &  2.6 & 2.2     & 44.8 & 11.29\\
                     & LS 5039              &  276.56 &  -14.83 & 0.26    &  2.1 & 4.0     & 52.3 &  2.72\\
Massive star cluster & HESS J1614-518       &  243.58 &  -51.82 & 0.0058  &  2.2 & 4.0     & 45.4 & 18.33\\
\enddata

\tablecomments{Please refer to \autoref{tab:sourcelist1} for comments regarding
               this table.}
\tablenotetext{a}{Upper limits are in units of
$\zeta=10^{-12}\flux$}
\tablenotetext{b}{Most significant source in this table (South only).
The trial-corrected p-value is $9.3\%$.}

\end{deluxetable*}

\end{document}